\documentclass[12pt,a4paper]{article}
\usepackage{a4wide}
\usepackage{slashed}
\usepackage{graphicx}
\def\La{\Lambda}
\def\bea{\begin{eqnarray}}
\def\eea{\end{eqnarray}}         
\def\beq{\begin{equation}}
\def\eeq{\end{equation}}
\def\e{\mathrm{e}}
\def\im{\Im\mathrm{m}}
\def\al{\alpha}
\def\be{\beta}
\def\ga{\gamma}
\def\de{\delta}
\def\ep{\epsilon}
\def\la{\lambda}
\def\si{\sigma}
\def\nn{\nonumber}
\def\dl{L_2}
\def\GG{\langle G^2\rangle}
\def\GGG{\langle G^3\rangle}
\def\aGG{\langle \alpha_s G^2\rangle}
\def\FB{F_{B_c}}
\def\MB{M_{B_c}}
\def\Mom#1{\mathcal{M}^{(#1)}}
\def\Lc{\mathcal{L}}
\def\R{{\mathcal{R}}}
\def\ps{\slashed{p}}
\def\barr{\begin{array}}
\def\ear{\end{array}}
\def\Pb{\bar{\psi}}
\def\alto{}

\def\rar{\rightarrow}
\begin{document}

\begin{titlepage}
\begin{flushright}
{\small
UAB-FT-325 \\
CERN-TH.7141/94 \\
HD-THEP-93-51 \\
ISN 93--123 \\
PM-93-44}
\end{flushright}
\begin{center}
{\Large\bf HADRONS WITH CHARM AND BEAUTY}
\vglue 0.2cm
{{\bf E.~Bagan$^{a),b),c)}$, H.G.~Dosch$^{b)}$, P.~Gosdzinsky$^{a)}$, \\
 S.~Narison$^{d),e)}$}
 and {\bf J.-M.~Richard}$^{f),g)}$}\\
\vskip 0.2cm
$^{a)}${\small Grup de F\'{\i}sica Te\`orica, Dept. de
F\'{\i}sica i Institut de F\'\i sica d'Altes Energies, IFAE,}\\
{\small Universitat Aut\`onoma de Barcelona},\\
{\small 08193 Bellaterra, Spain}\\
\vskip .1cm
$^{b)}${\small Institut f\"ur Theoretische Physik},
{\small Universit\"at Heidelberg, \\
 Philosophenweg 16},
{\small 69120 Heidelberg, Germany}\\
\vskip .1cm
$^{c)}${\small Alexander Von-Humboldt Fellow}\\
\vskip .1cm
$^{d)}${\small Laboratoire de Physique Math\'ematique},
{\small Universit\'e de Montpellier 2},\\
{\small 34095 Montpellier, France}\\
\vskip .1cm
$^{e)}${\small CERN, Theory Division},
{\small CH-1211 Gen\`eve 23, Switzerland},\\
\vskip .1cm
$^{f)}${\small Institut des Sciences Nucl\'eaires},
{\small Universit\'e Joseph Fourier--CNRS--IN2P3}\\
{\small 53, avenue des Martyrs, 38026 Grenoble, France}\\
\vskip .1cm
$^{g)}${\small European Centre for Theoretical Studies}
{\small in Nuclear Physics and Related Areas (ECT)}\\
{\small Villa Tambosi, strada delle Tabarelle, 286},
{\small 38059 Villazzano (Trento), Italy}\\
\end{center}
\vskip 0.30cm
{{\bf Abstract:} \small\rm By combining potential models
and QCD spectral sum rules (QSSR),
we discuss the spectroscopy
of the $(b\bar c)$ mesons and of the $(bcq)$, $(ccq)$ and $(bbq)$
baryons
(${q}\equiv {d}$ or $s$),
the
decay constant and the (semi)leptonic
decay modes of the $B_c$ meson.
For the masses, the best predictions come
{}from potential
models and read:
$M_{B_c} = (6255 \pm 20)$~MeV,
$M_{B^*_c} = (6330 \pm 20)$~MeV,
$M_{\Lambda(bcu)} = (6.93\pm 0.05)$~GeV,
$M_{\Omega(bcs)} = (7.00\pm 0.05)$~GeV,
$M_{\Xi^*(ccu)} = (3.63\pm 0.05)$~GeV and
$M_{\Xi^*(bbu)} = (10.21\pm 0.05)$~GeV.
The decay constant $f_{B_c}
= (2.94 \pm 0.21) f_\pi $ is well determined from QSSR and leads to:
$\Gamma(B_c \rightarrow \nu_\tau \tau) = (3.0 \pm 0.4)( V_{cb}/0.037
)^2$
  $\times 10^{10}$ s$^{-1}$.
 The uses
of the vertex sum rules for the semileptonic decays of the $B_c$ show
that the $t$-dependence of the form factors is much stronger than predicted
by vector meson dominance. It also
 predicts the almost equal strength of
  about 0.30
  $\times 10^{10}$ sec$^{-1}$
for the
semileptonic rates $B_c$ into $B_s, B^*_s,\eta_c$ and J/$\psi$.
Besides these phenomenological results, we also show explicitly how the
Wilson coefficients of the $\langle\alpha_s G^2\rangle$ and $\langle
G^3\rangle$ gluon condensates
already contain the full heavy quark- ($\langle\bar QQ\rangle$) and
mixed- ($\langle\bar QGQ\rangle$) condensate contributions in the OPE.}
\vskip 0.5cm
\noindent February, 1994
\end{titlepage}
\goodbreak
\section{Introduction}
\label{Intro}

With the planned high-energy machines such as the LHC, $B$-factories,
the Tevatron with high luminosity,
there is some hope and possibility
to identify and study hadrons containing two heavy quarks \cite{LHC},
like
double-charm baryons $(ccq)$ or hadrons with charm and beauty,
namely $(b\bar c$) mesons and $(bcq)$ baryons. Here, and
throughout this paper, $q$ denotes a light quark $u$ or $d$.\par
In view of this project, it is important to have  safe theoretical
predictions as a guide to the experimental searches of these hadrons.
There are already some theoretical studies on
($b\bar c$) states. To our knowledge, the pioneering works
on this analysis are the ones in Ref.\cite{Eich-Fein} from potential
models and the ones in Ref. \cite{SN1}
{}from QCD Spectral Sum Rules (QSSR) \` a la SVZ \cite{SVZ}.
In this
paper,
we are interested in the following topics.
\begin{itemize}
\item[i)] Masses: so far the ground-state masses of hadrons
exhibit nice regularities in flavor space, as illustrated by
the Gell-Mann--Okubo mass formula, the equal-spacing rule
of decuplet baryons, etc.; we would like to know the analogue
of these regularity patterns in the sector of heavy quarks, and
in particular interpolate $(b\bar c)$ from
$(c\bar c)$ and $(b\bar b)$, and extrapolate
{}from  single-charm $(cqq)$ and single-beauty $(bqq)$ baryons toward
$(bcq)$ baryons with both charm and beauty.
\item[ii)] Decay constants:  we know, in the case of the heavy--light
quark systems, that the decay constants of the $D$ and $B$ mesons
do not yet satisfy  the 1/$\sqrt{M_Q}$ heavy quark scaling due to
large 1/$M_Q$ corrections and that the prediction of the potential
models based on the meson wave function fails. Then,
we would like to test if the $B_c (\bar bc)$
meson decay constant can be predicted reliably from the potential models
by comparing it with the one from QSSR.
\item[iii)] Semileptonic decay properties: we also know that QSSR
vertex sum rules can predict successfully the semileptonic widths
of the $D$ and $B$ mesons. Then, we pursue this application in the case
of the $B_c$ meson.
\end{itemize}
It should be noted that the Heavy Quark Effective Theory (HQET)
\cite{MAN}, which is successful in the heavy--light quark systems,
cannot be applied straightforwardly to the $\bar bc$
and $\bar bcu$ states, unless the charm-quark mass is considered
to be light, which is not a good approximation. Therefore,
we combine the potential models with the QSSR
approaches for estimating the masses and/or couplings of the $\bar
bc$ and
$\La(bcu)$ states.
 The former is known to have successful predictions for the hadron
masses, while its connection
with QCD starts to be understood within the framework
of HQET. QSSR is also known to describe
successfully the hadron properties, although the accuracy of its
predictions for
the meson masses is limited by the systematic of the method and is less
than the potential model ones. In the other cases, such as the couplings and
decays, QSSR predictions are more precise and reliable.

The aim of this paper is twofold.
First, we summarize the rigorous results of potential
models such as mass
inequalities and bounds on short-range correlations. We also present typical
predictions for a ``realistic" phenomenological potentials for the
$\bar bc$ and
$\La(bcu)$ states.
Secondly, we present improved results for the masses, couplings and
form factors of the
semileptonic decays from the QSSR approach.

\section{Results of potential models}
\label{Pot-mod}
In this section,
we first give
brief reminders of general results from potential models,
which can be found in
reviews~\cite{QR1}--\cite{GMbook}, with
references to the original papers. We then
summarize  the
rigorous and empirical results of potential models: mass
inequalities, bounds on short-range correlations,
typical predictions for masses and decay constants focused
on the
applications to the particular  ($b\bar c)$  mesons and
$(bcq)$ baryons.\par
\subsection{Constraints on the $b\protect{\bar{c}}$ mass}
\label{Constr-bc}
Consider  a purely central and flavor-independent
potential. Then the binding energy depends on the flavor
of the constituents only through the inverse  masses
$m_1^{-1}$ and $m_2^{-1}$, which enter
the Hamiltonian linearly. At fixed $m_1$, the lowest energy is an
increasing and concave function of $m_2^{-1}$ \cite{Thir,BeMa}.
One can for instance extrapolate the $(b\bar c)$ energy out
of the ($b\bar s$) and $(b\bar q$) energies. This gives an upper
limit:
 \begin{equation}\label{ineg-bc-3}
E( b\bar c)\le E( b\bar s)
{m^{-1}_c-m^{-1}_q\over
m^{-1}_s -m^{-1}_q}+E( b\bar q)
{m^{-1}_c-m^{-1}_s\over
m^{-1}_q-m^{-1}_s}.
 \end{equation}
It is independent of the $b$-quark mass, but depends upon the
inverse quark masses $m^{-1}_c$, $m^{-1}_s $ and $m^{-1}_q$,
which are not directly observable. Anyhow, (\ref{ineg-bc-3}) is
not very accurate, since ($b\bar s$) and ($b\bar{
q}$) are too close to each other to allow for a precise
determination of the limiting straight line, in a plot of meson
energies versus the inverse constituent masses.

In fact, better results are obtained by separating out the
centre-of-mass motion, and using the inverse
reduced mass $\alpha=m_1^{-1}+m_2^{-1}$, which enters the
relative Hamiltonian linearly. The ground state is an increasing
and concave function of $\alpha$ \cite{Thir,BeMa}. Thus
\begin{equation}
\label{ineg-bc-1}
( b\bar c)\ge{(c\bar c)+( b\bar{b})\over2}.
\end{equation}

For numerical applications of (\ref{ineg-bc-1}), one has to
consider the spin-averaged masses, such as:
\begin{equation}
\label{meson-spin-ave1}
(c\bar c)=
{1\over4}\eta_c+{3\over4}{\rm J}\!/\Psi
\end{equation}
and its ($b\bar b$) analogue, with the  results
\begin{equation}
\label{meson-spin-ave2}
(c\bar c)=3.067\,\mbox{GeV}, \qquad( b\bar{b})=9.448\,\mbox{GeV}
\end{equation}
where experimental masses \cite{PDG} are used, and an hyperfine
splitting $\Upsilon-\eta_ b=50  $ MeV is assumed.
This gives a lower limit
\begin{equation}
\label{bc-low-num}
( b\bar c)\ge6.257\,\mbox{GeV}
\end{equation}
for the spin averaged ($b\bar c)$ state.

An upper limit is also obtained from the same concavity behavior
in the inverse reduced mass $\alpha$:
\begin{equation}
\label{ineg-bc-2}
( b\bar c)\le( b\bar s)+(c\bar{s})-(s \bar s).
\end{equation}
If one uses
\begin{equation}
\label{bc-up-input}
(c\bar s)=2075\,\mbox{MeV},\quad
( b\bar s)=5390\,\mbox{MeV},\quad
(s\bar s)=950\,\mbox{MeV},
\end{equation}
 one gets
\begin{equation}
\label{bc-up-num}
( b\bar c)\le 6.52\,\mbox{GeV}.
\end{equation}

We suspect that this bound is not very accurate, and therefore not too
reliable, because it involves the strange quark.
In fact one can derive an upper bound involving heavy quarks
only, provided one also accounts for the excitation spectrum.
The reasoning below is inspired by the work of Martin and
Bertlmann \cite{BeMa}.

{}From the Feynman--Hellmann theorem \cite{Thir},
\begin{equation}
\label{Feynmann-H}
{{\rm d}E\over{\rm
d}\alpha}=\langle{\rm\bf p}^2\rangle={T(\alpha)\over\alpha},
\end{equation}
where $T$ denotes the expectation value of the kinetic energy,
we have
\begin{equation}\label{bb-to-bc}
E( b\bar c)=E( b\bar b)+
\int_{\alpha( b\bar b)}^{\alpha( b\bar{c})}\alpha^{-1}T(\alpha){\rm d}\alpha .
\end{equation}
and
\begin{equation}
\label{cc-to-bc1}
E( b\bar c)=E(c\bar c)-\int_{\alpha(b\bar c)}^{\alpha(c\bar c)}
\alpha^{-1}T(\alpha){\rm d}\alpha
\end{equation}
We now make the mild restriction that the potential $V$ is
intermediate between Cou\-lomb and linear, and {\sl a fortiori}
intermediate between Coulomb and harmonic. More precisely, we
assume $\Delta V\ge 0$ and $V''\le 0$. Then 

\noindent
{\it i)} $T(\alpha)$ is intermediate between $\alpha^{-1}$
(Coulomb) and $\alpha^{1/3}$ (linear), i.e.\ $\alpha T(\alpha)$
increases with $\alpha$ while $\alpha^{-1/3}T(\alpha)$ decreases;

\noindent
{\it ii)} if
$\delta E=\left[E_{\rm 1P}(\alpha)-E_{\rm 1S}(\alpha)\right]/4$
denotes the orbital excitation energy,
the ratio $T/\delta E$ is larger than $3/4$ (harmonic) and smaller
than $4/3$ (Coulomb).\par
After some manipulations, we obtain
\begin{eqnarray}
\label{bc-upper-1}
&&M( b\bar c)\le
 M(c\bar c)+(m_ b-m_c)
 -{9\over4}\delta E(c\bar c)\left[1-
\left({m_ b+m_c\over 2m_{
b}}\right)^{1/3}\right]\nonumber\\
&&M( b\bar c)\le
 M( b\bar b)-(m_ b-m_c)
 +4\delta E( b\bar b)\left[
\left({m_ b+m_c\over 2m_{
c}}\right)^{1/3}-1\right]\nonumber . \\
\end{eqnarray}
 When they are combined,
most of the dependence on the constituent masses disappears, and
we obtain:
\begin{equation}
\label{bc-upper2}
( b\bar c)\le{(c\bar c)+( b\bar{
b})\over2}-{9\over8}\delta E(c\bar c)\left[1-
\left({m_ b+m_c\over 2m_{
b}}\right)^{1/3}\right]+2\delta E( b\bar b)\left[
\left({m_ b+m_c\over 2m_{
c}}\right)^{1/3}-1\right]
\end{equation}
After proper spin averaging of the orbital excitations \cite{PDG},
one finds:
 $\delta E( b\bar b)\simeq
\delta E(c\bar c)\simeq0.45\,$GeV.
If one takes $m_ b/m_{
c}=3$, one  obtains
\begin{equation}
\label{bc-upper3}
( b\bar c)\le6.43\,\mbox{GeV}.
\end{equation}

Instead of working with spin-averaged masses, one could in
principle write inequalities relating   pseudoscalar
states. If, indeed, the additional term
is (including $\vec\sigma_i\cdot\vec\sigma_j=-3$) of the form
\begin{equation}
\label{spin-spin}
\delta V=-{3\over m_1m_2}V_{\rm SS},\qquad V_{\rm SS}>0,
\end{equation}
then the whole Hamiltonian is a linear function of $m_1^{-1}$ at
fixed $m_2^{-1}$, or a concave function of $m^{-1}$
for $m_1=m_2=m$,
and one can still write some convexity inequalities. The problem
is the lack of accurate experimental input for the pseudoscalar
masses.

\subsection{Explicit calculations of $b\protect{\bar{c} }$
ground state}
\label{Explicit-bc}
To  estimate  the departure from a simple additive ansatz
$2( b\bar c)
=(c\bar c)+( b\bar b)$,
one can use a logarithmic potential, which is
known as a good approximation to more elaborate potentials
\cite{QR1}. If $V=A+B\ln(r)$, then the ground-state energy is of
the form $E=A'-B\ln(\mu)/2$. With typically $m_ b/m_{c}=3$
 and $B\sim0.7\,$ GeV, one gets an effect
\begin{equation}
\label{bc-log}
( b\bar c)-{(c\bar c)+( b\bar{
b})\over2}\simeq0.1\,\mbox{GeV},
\end{equation}
which is of course compatible with the inequalities written in the
previous section.

Let us now collect some predictions of typical potential
models proposed in the literature.
In Ref.\ \cite{ccc2}, A.\ Martin applied to (b$\bar c$)
his simple power-law potential. It consists of
\begin{equation}\label{Martin-fit1}
V=A+B r^{0.1}+
C{\vec\sigma_1\cdot\vec\sigma_2\over m_1m_2}
\delta^{(3)}({\rm\bf r}_2-{\rm\bf r}_1),
\end{equation}
with $A=8.064$, $B=6.870$ and $C=1.172$,
in units of powers of GeV.
The quark masses are constituent masses and are
$m_s=0.518\,$GeV,
$m_c=1.8\,$GeV,
and $m_ b=5.174\,$GeV.  The spin--spin term is treated  at
first order. It is adjusted to reproduce
the J/$\!\Psi-\eta_c$ mass splitting (112 MeV \cite{PDG}).
He obtained
\begin{equation}\label{Martin-fit2}
{ b\bar c}(0^-)=6.25\,\mbox{GeV}\qquad
{ b\bar c}(1^-)=6.32\,\mbox{GeV},
\end{equation}
corresponding to an average of $6.30\,$GeV.
These are the values also obtained by Gershtein et al.\
\cite{Gershtein}, who used essentially the same potential.
Previously, Eichten and Feinberg, in the course of their study
of spin-dependent forces \cite{Eich-Fein}, considered the
$(b\bar c$) system, and got
\begin{equation}\label{Eich-Fein-bc}
{ b\bar c}(0^-)=6.24\,\mbox{GeV}\qquad
{ b\bar c}(1^-)=6.34\,\mbox{GeV}.
\end{equation}
More recently, Eichten and Quigg \cite{Eichten-Quigg-bc}
estimated
\begin{equation}\label{Eich-Quigg-bc}
{ b\bar c}(0^-)=6.26\,\mbox{GeV}\qquad
{ b\bar c}(1^-)=6.33\,\mbox{GeV},
\end{equation}
with a typical uncertainty of $\pm 20\,$MeV, from a survey of
realistic quarkonium potentials.
 \par
One can go a little beyond the frame of this section and look
at constituent models with relativistic forms of kinetic energy.
They lead to the same kind of regularities as non-relativistic
models, although the corresponding theorems are not always
available in a fully rigorous and general form. For instance,
Goodfrey and Isgur obtained
\begin{equation}
{ b\bar c}(0^-)=6.27\,\mbox{GeV}\qquad
{ b\bar c}(1^-)=6.34\,\mbox{GeV},
\end{equation}
in their model \cite{God-Isg}, which tentatively describes
all mesons, light or  heavy.\par
As often in this field, there is a nice convergence of
all potential models, and the uncertainty of $\pm 20\,$MeV
estimated by Eichten and Quigg seems rather safe. By taking the
average of different estimates and by adopting the previous uncertainty,
we obtain the final estimate:
\begin{equation}
{ b\bar c}(0^-)=(6255 \pm 20)\,\mbox{MeV}\qquad
{ b\bar c}(1^-)=(6330\pm 20)\,\mbox{MeV}.
\end{equation}

\subsection{Decay constant of mesons}

For the estimate of the decay  constants, let us consider the meson
wave function:
\begin{equation}
\label{def-p}
p=\left\vert\Phi(0)\right\vert^2=\langle\Phi\vert
\delta^{(3)}({\rm\bf r}_2-{\rm\bf r}_1)\vert\Phi\rangle,
\end{equation}
which governs the leptonic widths, hadronic widths, etc.
It also enters the calculation of hyperfine splittings, when a simple
contact term as that in Eq.\  (\ref{Martin-fit1}) is adopted.

To estimate how $p$ varies from one meson to another, let us
consider first a power-law potential $V\propto r^\beta$.
Then, from the well-known scaling laws \cite{QR1,JMRrep}, one
gets
 \begin{equation}
\label{scaling-p}
p(\alpha)\propto \alpha^{3/(\beta+2)},
\end{equation}
as a function of the inverse reduced mass $\alpha$. In particular,
one expects $ p\propto \alpha^{2/3}$ for a logarithmic potential,
which is known to mimic the good potentials in the region of
interest.

Note that one cannot object that, $p$ being the square wave
function at zero separation, it is extremely sensitive to the
very short-range part of the potential. In fact $p$ is given
by the potential in the region where the wave function is
important. This is seen on the so-called
Schwinger rule \cite{QR1}
\begin{equation}
\label{Schwinger}
p={1\over4\pi\alpha}\int{\rm dr}^{(3)}\vert\Phi(r)\vert^2{{\rm
d}V\over{\rm d}r}.
\end{equation}
In short, we expect  regular increases of $p$ when one goes from
$c\bar c$ to $b\bar b$ via $b\bar c$, and
presumably
\begin{equation}
\label{ineg-p}
p( b\bar c)\le{1\over2}\left[p( b\bar b)
+p(c\bar c)\right].
\end{equation}

If one uses the potential model of Eq.\ (\ref{Martin-fit1}),
one obtains, in units of GeV$^3$ :
\begin{equation}
\label{p-Martin}
p(c\bar c)=0.077, \qquad
p( b\bar b)=0.350, \qquad
p( b\bar c)=0.136.
\end{equation}
The absolute values are less reliable than the relative ones.
Similarly, potential models usually fail in predicting the
leptonic widths of the $J/\Psi$ and its radial excitations, or
of the $\Upsilon$ states, but give a fair account of the
ratios of leptonic widths. In terms of the wave function, the decay
constant reads :
\begin{equation}
f_P=\sqrt{\frac{6p}{M_P}},
\end{equation}
while its normalization in terms of the quark currents is:
\bea
(m_c+m_b)<0|\bar c(i\gamma_5)b|B>=\sqrt{2}M^2_{B_c}f_{B_c} \nonumber \\
<0|\bar b\gamma^\mu b|\Upsilon>=\sqrt{2}M_{\Upsilon}f_{\Upsilon}
\epsilon^\mu .
\eea
Then, we deduce from (27):
\beq
f_{B_c} \simeq (3.86\pm 1.31)f_\pi.
\eeq
The error in this result comes from the departures of different
potential-model
predictions \cite{Gershtein}, \cite{LUSI},
\cite{Eichten-Quigg-bc} from our value. It will be
compared in section 3 with the QSSR estimates.

\subsection{Inequalities on baryon masses}
\label{Ineg-baryons}
Let us start with a flavor- and spin-independent potential
$V(\vec r_1,\vec r_2,\vec r_3)$.
\par
For every potential $V$, the ground-state energy is a concave
function of each inverse mass $m_i^{-1}$. One could for instance
set an upper limit on $(bcq)$ in terms of $(ccq)$ and $(csq)$, or in
terms of $(csq)$ and $(cqq)$, and the corresponding quark masses.
Again, it is not very useful to write inequalities that involve
unobservable quark masses.\par
With mild restrictions on the shape of the potential, one can
write convexity relations in terms of actual hadron masses
\cite{Liebetc}. For instance, there is a generalization of
(\ref{ineg-bc-1})
\begin{equation}\label{ineg-bcq-1}
(bcq)\ge{(bbq)+(ccq)\over2},
\end{equation}
or the even more exotic looking \cite{Martin-bcs}
\begin{equation}\label{ineg-bcq-2}
(bcq)\ge{( bbb)+( ccc)+( qqq)\over3}.
\end{equation}
For numerical applications
with the presently available data, one would prefer the generalization
of (\ref{ineg-bc-2})
\begin{equation}
\label{ineg-bcq-3}
(bcq)\ge(bqq)+(cqq)-(qqq).
\end{equation}
This gives as a rough estimate
\begin{equation}
\label{ineg-bcq-4}
(bcq)\ge 6.9\,\mbox{GeV},
\end{equation}
if one uses the rounded values
$m(bqq)=5.6$, $m(cqq)=2.4$, and $m(qqq)=1.1\,$GeV.

\subsection{Relations between mesons and
baryons}
\label{mes-bar}
We suppose here that there is a simple relation between
the potentials governing mesons and baryons:
\begin{equation}
\label{half}
V(\vec r_1,\vec r_2,\vec r_3)={1\over2}\sum_{i<j}V_{
q\bar q}(|\vec r_i-\vec r_j|).
\end{equation}
There is no profound justification for this rule in QCD.
We simply remark that it seems compatible with the present
phenomenology. In particular, it leads to amazing inequalities
among meson and baryon masses \cite{JMRrep}. These inequalities
are always satisfied when they can be checked, so one is tempted
to believe that they can also hold for baryons that have not  yet been
discovered. For instance,
 \begin{equation}
\label{meson-baryon-1}
(bcq)\ge{( b\bar c)+( b\bar q)+(c\bar q)\over 2}.
\end{equation}
With the spin-averaged masses $m(B)=5.3$ and $m(D)=1.97\,$GeV,
and with our previous lower bound (\ref{bc-low-num}) on ($b\bar
c$), one obtains
 \begin{equation}
\label{meson-baryon-2}
(bcq)\ge 6.73\,\mbox{GeV}.
\end{equation}
We suspect this to be a rather crude lower bound,
and, indeed, it does not improve our previous lower
bound (\ref{ineg-bcq-4}). In
deriving Eq.\ (\ref{meson-baryon-1}), one neglects the motion of
the  centre of mass of any quark pair in the overall rest frame
of the baryon. Improvements are feasible, to better express
3-body energies in terms of 2-body energies, but the latter are
no longer too easily expressed as energies of actual mesons
\cite{Post,BMR2}.
\subsection{Explicit model calculations of $(bcq)$ masses}
Unfortunately, there are not too many explicit computations
of the masses of baryons with two heavy quarks, at least to our
knowledge. The case of $(ccq)$ baryons was considered by Fleck and
Richard \cite{FR1}. They first use a non-relativistic potential
model. Not surprisingly, the exact solution of the 3-body
problem is well reproduced by a Born--Oppenheimer approximation.
This opens the possibility of treating the light quark
relativistically, for a fixed separation of the heavy quarks.
This was done in Ref.\ \cite{FR1}, where a variant of the
MIT\ bag model was used. It was found, however, that
the  results are rather sensitive to the details of the bag
model. We shall not consider them further and restrict ourselves
to the potential-model picture. In principle, the
Born--Oppenheimer treatment could be repeated, with the gluon
and light-quark degrees of freedom treated via sum rules or via
a lattice simulation, at fixed QQ separation.

The results for (ccq) are obtained with a
simple local and pairwise interaction
\begin{equation}
\label{pot-bcq}
V_{\rm T}={1\over2}\sum_{i<j}V,
\end{equation}
where the factor $1/2$ is an arbitrary convention (though
reminiscent from the discussion in Sec.\ \ref{mes-bar}, and $V$ is a
variant of the power-law potential (\ref{Martin-fit1}), adjusted
to fit all ground-state baryons \cite{RiTa2}. The parameters are
$A=-8.337$, $B=6.9923$, $C=2.572$, where units are  powers of GeV.
 As for the constituent masses, which should not be confused
with the masses used in the QSSR analysis, we use
$m_{\rm q}=0.300$,
$m_{\rm s}=0.600$, and $m_{\rm c}=1.905\,$GeV.
The latter value is 10 MeV above the c-quark mass
 in Refs.\ \cite{RiTa2,FR1}, to better reproduce the
experimental mass of the $\Lambda_{\rm c}$ at 2285 MeV \cite{PDG}.
The $\Sigma_{\rm c}-\Lambda_{\rm c}$ difference comes out right.
If one takes for the b quark a mass $m_{\rm b}=5.290$,
one obtains a reasonable $\Lambda_{\rm b}$ at $5.620\,$GeV, which is
the central value recently reported \cite{Lambdab}.

We keep these parameters fixed to calculate the masses
given in Table~\ref{table1}, namely
the spin-averaged mass $\,\overline{\!M}$ (computed without the
spin--spin term), and the lowest spin-$1/2$ state.
\begin{table}
\begin{center}
\begin{tabular}{ccc}
State&$\,\overline{\!M}$&$M_0$\\
\hline
$(ccq)$&\phantom{1}3.70&\phantom{1}3.63\\
$(ccs)$&\phantom{1}3.80&\phantom{1}3.72\\
$(bbq)$&10.24&10.21\\
$(bbs)$&10.30&10.27\\
$(bcq)$&\phantom{1}6.99&\phantom{1}6.93\\
$(bcs)$&\phantom{1}7.07&\phantom{1}7.00\\
\hline
\end{tabular}
\end{center}
\caption{\label{table1} Masses of heavy baryons with a simple
power-law potential fitted to known baryons.
$\,\overline{\!M}$ is the spin averaged mass, and $M_0$ that of
the lowest state, with spin-$1/2$. Units are GeV.}
\end{table}

A remark concerning the spin structure: the lowest $(ccq)$ baryon
has spin $S=1/2$, with the $(cc)$ pair in a spin $s=1$ state, as
dictated by the statistics. For $(bcq)$, we have a mixing of $s=0$
and $s=1$, with the latter dominating, to leave maximal strength
for $(qc)$ and $(qb)$ pairs (for total spin $S=1/2$,
the cumulated $\sum_{i<j}
\vec\sigma_i\cdot\vec\sigma_j$ is fixed at the value $-3$,
independent of the internal spin structure).

We estimate the theoretical uncertainty around $\pm 20$ MeV
in the extrapolation. The main additional
uncertainty comes from the mass of $\Lambda_{\rm b}$. Altogether
we obtain
\begin{equation}
\label{final-baryons}
\Lambda(bcq)=6.93\pm0.05\,\mbox{GeV}\quad
\Omega(bcs)=7.00\pm0.05\,\mbox{GeV}.
\end{equation}
We can also deduce from Table 1, the masses of the ${\Xi^*_c}(ccu)$
and ${\Xi^*_b}(bbu)$ with the same degree of accuracy of 50 meV. The
result for $\Lambda(bcq)$ agrees quite well with the improved QSSR
estimate which will be discussed in the next section. The ones for
$\Xi^*_{c,b}$
 agree  with the QSSR estimates in \cite{CHAB4} which will also be
reminded section 3.
\subsection{Short-range correlations in baryons}
\label{p-baryons}
The quantity $p$ defined in Eq.\ (\ref{def-p}) for mesons is
generalized as
\begin{equation}
\label{def-pij}
p_{ij}=\langle\Phi\vert
\delta^{(3)}({\rm\bf r}_j-{\rm\bf r}_i)\vert\Phi\rangle.
\end{equation}
We are not aware of too many results on the coefficients
$p_{ij}$. The Schwinger rule (\ref{Schwinger}) has been
generalized \cite{HSF}, but the sum rule now involves
centrifugal barriers (in an $s$-wave baryon, the pairs are not
strictly in a  state of orbital momentum $\ell=0$, except
in the harmonic-oscillator case), and angular correlations like $\hat {\rm
r}_{ij}\cdot \hat {\rm r}_{ik}$. The available results concern
symmetric and nearly symmetric cases. References can be found in
\cite{JMRrep}.

For the very asymmetric cases we are dealing with, we
simply read the values of the $p_{ij}$ from the wave function,
which is computed with our
simple power-law potential, using the method of hyperspherical
harmonics \cite{JMRrep}. The results are shown in Table~\ref{table2}.
\begin{table}
\begin{center}
\begin{tabular}{ccccc}
State&$p_{12}$&$p_{23}$&$p_{31}$&$10^3p_{123}$\\
\hline
$(ccq)$&0.039&0.009&0.009&0.36\\
$(ccs)$&0.042&0.019&0.019&0.36\\
$(bbq)$&0.152&0.012&0.012&4.08\\
$(bbs)$&0.162&0.028&0.028&4.09\\
$(bcq)$&0.065&0.010&0.011&0.90\\
$(bcs)$&0.071&0.021&0.025&0.90\\
\hline
\end{tabular}
\end{center}
\caption{\label{table2} Short-range correlation coefficients
$p_{ij}$ calculated with our simple power-law potential. Units are
GeV$^{3}$ for the 2-body terms $p_{ij}$, and GeV$^{6}$ for $p_{123}$.}
\end{table}
Some remarks are in order:

\noindent
{\sl i)} The correlation between two quarks depends on the third one
\cite{CohLip}.

\noindent
{\sl ii)} There are more correlations between $b$ and $\bar c$ in
a ($b\bar c$) meson than between $b$ and $c$ in $(bcq)$ or $(bcs)$.

The coupling constants $\vert Z\vert^2$ that are usually quoted (see, e.g.
Ref. \cite{CHAB1}--\cite{CHAB3}, \cite{CHAB4})
have more to do with the probability $p_{123}$ of finding the
three quarks at the same place in the non-relativistic wave function.
Some values of $p_{123}$ are shown in Table \ref{table2}.
The normalization requires some technicalities. We define
\begin{equation}
\label{def-p123}
p_{123}=\langle\Phi\vert
\delta^{(3)}({\rm\bf x})\delta^{(3)}({\rm\bf y})\vert\Phi\rangle,
\end{equation}
where the Jacobi variables are introduced as
\begin{eqnarray}
\label{Jacobi}
&&{\rm\bf x}={\rm\bf r}_2-{\rm\bf r}_1\nonumber\\
&&{\rm\bf y}=\left[{\rm\bf r}_3-{ m_1{\rm\bf r}_1+m_2{\rm\bf r}_2
\over m_1+m_2}\right](m_1+m_2)\sqrt{ {m_3\over m_1m_2(m_1+m_2+m_3)} }
\end{eqnarray}
(the coefficient of ${\rm\bf y}$ is such
that the kinetic energy operator is proportional to
${\rm d}^2/{\rm d}{\rm\bf x}^2+{\rm d}^2/{\rm d}{\rm\bf y}^2$),
and the labeling is such that 1 and 2 are the heavy quarks,
and 3 the light one.



\section{The $B_c$, $\La(bcu)$ , ${\Xi^*_c}(ccu)$ and ${\Xi^*_b}(bbu)$
masses and couplings from QSSR}
\label{S3}

We have studied in the previous section the properties of the $B_c$
meson, $\Lambda(bcq)$, $\Xi^*_c$ and $\Xi^*_b$
baryons using potential models. In the following,
we shall study their properties using the QSSR approach.

\subsection{The $B_c$-meson correlator}
\label{SS31}

We shall be concerned with the two-point correlator:
\beq
\psi_5(q^2)=i\int d^4 x\,\e^{iq\cdot x}
\langle 0| {\bf T} J_5(x) {J^{\dagger}}_{5}(0)|0\rangle
\label{3.1a}
\eeq
associated to the pseudoscalar current:
\beq
J_5(x)=(m_c+m_b) :\overline{b}(i\ga_5)c:
\label{3.1b}
\eeq

The spectral function $\im \psi_5(t)$ can be evaluated in QCD for $t\gg
\La^2$.
Its perturbative part is known to two loops in terms
of the pole quark masses~\cite{BG1}. It reads:
\bea
\mbox{Im} \psi_5^{\rm pert}(t) &=&
{3(m_b+m_c)^2\over8\pi t} \bar q{}^4 v \Bigg\{
1+{4\al_s\over3\pi} \Bigg\{ {3\over8}(7-v^2)\nn\\
&+&\sum_{i=b,c}\Big[ (v+v^{-1})\left(
\dl(\al_1\al_2)-\dl(-\al_i)-\log \al_1 \;\log \be_i
\right)
\label{N1}
\\
&+&
A_i \log \al_i +B_i \log \be_i \Big]\Bigg\}+O(\al_s^2)
\Bigg\}
\nn
\eea
where
\beq
\dl(x)=-\int_0^x {dy\over y} \log(1-y)
\eeq
and
\bea
A_i&=&{3\over 4}{3m_i+m_j\over m_i+m_j}-{19+2v^2+3v^4\over 32v}-
{m_i(m_i-m_j)\over
\bar q{}^2v(1+v)}\left(1+v+{2v\over 1+\al_i}\right);\nn\\
B_i&=&2+2{m_i^2-m_j^2\over \bar q{}^2 v};
\label{N2}
\\
\al_i&=&{m_i\over m_j}{1-v\over
1+v};\qquad \be_i=\sqrt{1+\al_i}\;{(1+v)^2
\over 4v}\nn\\
\bar q{}^2&=&t-(m_b-m_c)^2;\qquad v=\sqrt{1-4{m_b m_c\over
\bar q{}^2}}
\nn
\eea

The non-perturbative pieces of Im $\psi_5(t)$ can
be introduced using an OPE {\em \`a la} SVZ~\cite{SVZ}. We shall
consider the contributions of operators up to dimension six.
Following the usual procedure in Ref.~\cite{NOV}, we obtain the Wilson
coefficients of the $\GG$ and $\GGG$ gluon condensates. The diagrams
involved
are shown in Fig.~1.
\begin{figure}[hb!]
 \centering
 \includegraphics[width=.5\textwidth]{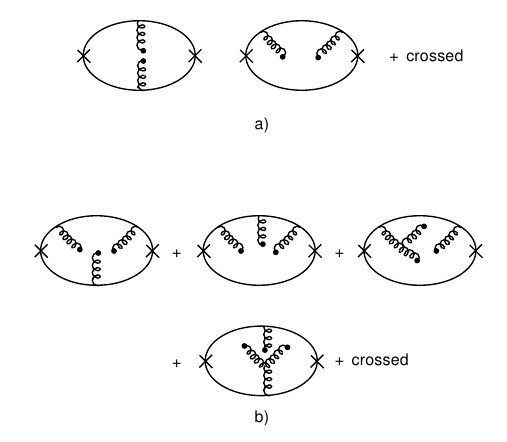}
 \caption{{\bf a)} Diagrams contributing to the
gluon condensate coefficient ${\rm Im}\, C_{G^2}$. {\bf b)}
Diagrams
contributing to the three-gluon condensate coefficient ${\rm Im}\,
C_{G^3}$.}
 \label{fig:fig1}
\end{figure}
Our results are:
\bea
\mbox{Im}\, C_{G^2}&=&-{\alpha_s  m_b m_c \,t
\over 2[t-(m_b-m_c)^2]^{3/2}}\nonumber\\
&\times&
\left( t-m_b^2-m_bm_c-m_c^2\right){\theta[t-(m_b+m_c)^2]
\over [t-(m_b+m_c)^2]^{5/2}}+\cdots
\label{3.3}\\
\mbox{Im}\,C_{G^3}&=&{\alpha_s m_b m_c\; t\over 6\;
[t-(m_b-m_c)^2]^{7/2}}
\left\{   3t^4
 -2(3m_b^2+2m_bm_c+3m_c^2)t^3  \right.\label{e}\nonumber \\
&+&  (5m_b^3m_c+18m_b^2m_c^2+5m_bm_c^3)\;t^2\nonumber\\
&+&   2(3m_b^6+m_b^5m_c-6m_b^4m_c^2-6m_b^3m_c^3-6m_b^2m_c^4+
m_bm_c^5+3m_c^6)\;t     \nonumber\\
&-& \left.   3(m_b^8+m_b^7m_c-m_b^5m_c^3-2m_b^4m_c^4-m_b^3m_c^5+m_b
m_c^7+m_c^8)
     \right\}\nonumber\\
&\times&{\theta[t-(m_b+m_c)^2]\over [t-(m_b+m_c)^2]^{9/2}}+\cdots
\label{3.4}
\eea
The dots in~(\ref{3.3}) and~(\ref{3.4}) stand for terms proportional to
$\de(t-(m_b+m_c)^2)$
and derivatives. They should be
there to compensate for the singular behavior (at threshold)
of $\im
C_{G^2}$ and $\im C_{G^3}$  in a dispersion
relation such as~(\ref{Ab}) in the appendix. One can
circumvent the problem of computing these terms by using the
method explained in the appendix.
Our result for $C_{G^2}$ (see~(\ref{Ad}) in the appendix) agrees with previous
ones~\cite{SN2}, while the one for $C_{G^3}$ is new. In the equal-mass case, it
agrees with the result in Refs.~\cite{C_G^3,Latorre}.

It should be emphasized that~(\ref{3.3}) and~(\ref{3.4}) already contain the
contributions of the $\langle \overline{c}c\rangle$ and $\langle \overline{c}G
c\rangle$  condensates through the heavy-quark expansion (see~(\ref{b})
and~(\ref{c})
below).
In order to prove this result,
let us compute $C_{G^2}$ and $C_{G^3}$ (obtained as in the appendix) for small
values of $m_c$, retaining only the singular pieces as $m_c\to0$:
\begin{eqnarray}
C_{G^2}\!\!\!\!&=&\!\!\!\! -{\alpha_s m_b\over12\pi(q^2-m_b^2) m_c}
-{\alpha_s m_b\, q^2\over4\pi (q^2-m_b^2)^3}\, m_c \log m_c^2
+\cdots
\label{-a'}\\
C_{G^3}\!\!\!\!&=&\!\!\!\!-{\alpha_s m_b\over360\pi(q^2\!\!-\!m_b^2) m_c^3}
+{\alpha_s (q^2-2 m_b^2)\over720\pi(q^2\!\!-\!m_b^2)^2 m_c^2}
+{\alpha_s m_b(15q^2\!\!-\! m_b^2)\over360\pi(q^2\!\!-\!m_b^2)^3 m_c}\!
+\!\cdots
\label{a}
\end{eqnarray}
We now show that the terms of $C_{G^2}$
and $C_{G^3}$ in~(\ref{-a'}) and~(\ref{a}) appear
because of the {\em heavy-quark expansion}, namely:
\begin{eqnarray}
\langle \overline{c}c \rangle&=& -{1\over12 m_c}{\alpha_s\over\pi}\langle
G^2\rangle-{1\over360m_c^3}{\alpha_s\over\pi}
\langle G^3\rangle+\cdots
\label{b}\\
\langle\overline{c}G c\rangle&=&{m_c\over2}\log m_c^2
\;{\alpha_s\over\pi}\langle G^2\rangle-{1\over12m_c}{\alpha_s\over\pi}
\langle G^3\rangle+\cdots
\label{c}
\end{eqnarray}
To see this, let us give the quark and mixed condensate coefficients
for the pseudoscalar current (which can be
found in~\cite{Jamin}, appendix~A).
In
our notation:
\begin{eqnarray}
C_{\bar{c}c}&=&{m_b\over q^2-m_b^2}+{2m_b^2-q^2\over2(q^2-m_b^2)^2}m_c
+{m_b^3 \over (q^2-m_b^2 )^3}m_c^2+\cdots \label{d}\\
C_{\bar{c}Gc}&=&-{m_b q^2\over2(q^2-m_b^2)^3}+\cdots \label{ee}
\end{eqnarray}
Note that multiplying~(\ref{d}) and~(\ref{ee}) by~(\ref{b})
and~(\ref{c}),
respectively, and
adding the two contributions, one obtains~(\ref{-a'}) and~(\ref{a}).
This clearly shows that
our results for $C_{G^2}$ and $C_{G^3}$ already contain the parametrization
of the quark
and mixed
 condensates in terms of purely gluonic operators,
  as already shown in the
literature (see for instance~\cite{BrGe}).

\subsection{The $B_c$-meson coupling}
\label{SS32}

The $B_c$-meson is introduced via its coupling $\FB$ as:
\beq
\langle0|J_5|B_c\rangle=\sqrt{2} \FB\MB^2,
\label{3.10}
\eeq
while the contribution of higher radial excited states are averaged from the
QCD continuum above the threshold $t_c$. After transferring the continuum
effect into the QCD side of the spectral function,
the coupling $\FB$ can be estimated
{}from the finite energy sum rule moments:
\beq
\Mom n =\int_{(m_b+m_c)^2}^{t_c}{dt\over t^{n+2}}{1\over \pi}
\mbox{Im}\, \psi_5(t)
\label{3.11}
\eeq
or the Laplace sum rule:
\beq
\Lc=\int_{(m_b+m_c)^2}^{t_c}dt\;\e^{-t\tau}{1\over \pi}
\mbox{Im}\, \psi_5(t),
\label{3.12}
\eeq
while the $B_c$-mass squared can be obtained from the ratios:
\bea
\R&=&{\Mom n \over \Mom n+1 }
\label{3.13}\\
\R_{\Lc} &=&-{1\over\Lc}{d\Lc\over d\tau}.
\label{3.14}
\eea
Here
$n$, $\tau$ and $t_c$ are in general free external parameters in the analysis,
so that the optimal results should be insensitive to their values (stability
criteria). The first QSSR estimates of the $B_c$-meson mass and
couplings~\cite{SN1} are:
\beq
M_{B_c}= (6.5 \pm 0.4)\, \mbox{GeV} \ \ , \ \
f_{B_c}= (3.7 \pm 0.5)f_\pi,
\eeq
where the  uncertainties due to the mass and to the subtraction scale
(This scale does not appear
in the present paper, as can be inferred from Refs.~\cite{BrGe,Latorre}.
Thus, the parametrization
given by Ref.~\cite{NSV} and used in the previous paper is not correct.)
entering in the mixed condensate
imply a large error in the
estimate of the coupling $\FB$. For improving this result,
we shall use the potential-model
predictions in eq.~(22) and estimate $\FB$ from the sum rules
in~(\ref{3.11}) and~(\ref{3.12}). We show the results of the analysis in
Fig.~2.

\begin{figure}[hb!]
 \centering
 \includegraphics[width=.5\textwidth]{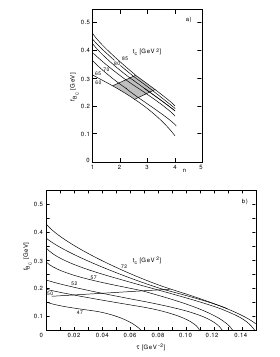}
 \caption{{\bf a)}$n$-dependence of the decay constant $f_{B_c}$ for
different values of the continuum threshold $t_c$.
{\bf b)} $\tau$-dependence of the decay constant $f_{B_c}$ for
different values of the continuum threshold $t_c$.}
 \label{fig:fig2}
\end{figure}

As one can see in this figure, the stability corresponds
to the inflexion point so that its localization is less precise than for
the case of the minimum (these inflexion points are indicated by the
shaded region in Fig. 2a and by the line in Fig. 2b).
 We assume that this will imply a $10\%$ error.
Taking the largest range of $t_c$-values from the onset of the
$n$- or $\tau$-stability region
($t_c \simeq 50$ GeV$^2$)
until the onset of the $t_c$-stability region
($t_c \simeq 67$ GeV$^2$)
and by taking the average of these two extreme values, we obtain:
\beq
\FB|_{Laplace} \simeq (2.95 \pm 0.27) f_\pi
\label{3.15}
\eeq
and:
\beq
\FB|_{Moments} \simeq (2.84 \pm 0.38) f_\pi.
\label{3.15'}
\eeq
We have used the values~\cite{SN2}:
\bea
\aGG&=&(0.06\pm0.02) \mbox{GeV}^4\nn\\
m_b(p^2=m_b^2)&=&(4.60\pm0.05)\mbox{GeV}
\label{3.16}
\\
m_c(p^2=m_c^2)&=&(1.47\pm0.05)\mbox{GeV}\nn\\
\langle g^3G^3\rangle&=&(1.2\mbox{GeV}^2)\times\aGG,
\nn
\eea
{}from a global
 QSSR analysis of different
  hadronic channels. The $\GGG$ value is based on a
rough estimate within the dilute gas instanton model~\cite{NSV}.

The main errors in $\FB$ come from the localization of the inflexion
point.
 One should notice that, at the inflexion point, the
$\al_s$-correction does not exceed $10\%$ of the leading-order term
for the two-point correlator. Contrary to the other QSSR analysis,
the non-perturbative terms are negligible and do not play a role in the
optimization procedure so that the optimal region is not well indicated.
However, the smallness of the non-perturbative terms
indicates that the OPE converges quite well at the optimization scale.
This value of $\FB$ agrees and improves (from the inclusion of the $G^3$-term)
the pioneer results in Refs.~\cite{SN1,SN2,SN3}.
Taking the average of the two QSSR values,
we deduce:
\beq
\FB|_{average} \simeq (2.94 \pm 0.22) f_\pi.
\label{3.15''}
\eeq
It is important to notice that the continuum energy $E_c$ defined as:
\beq
t_c \equiv (m_b+m_c+E_c)^2
\eeq
is:
\beq
E_c \simeq (1.0\sim 2.1)\, \mbox{GeV},
\eeq
in good agreement with what we know in the optimization of the sum
rule for the
heavy--light quark systems
\cite{SN4,ZAL}.
The result $\FB \simeq 1.22f_\pi$ obtained in
Ref.\cite{DOM} is too low, which should be due to numerical errors
as far as the result obtained from the moments in that paper
is concerned. The other possible source of uncertainties, in this paper,
 is the
value of the continuum threshold used in the analysis, which is too
low. The
result of Ref. \cite{COLA} is more similar to ours, but the procedure
used by the authors to derive it is very doubtful. Indeed,
we do not see any physical
reasons to move the $t_c$ values inside a small range
{}from 47 to 50 GeV$^2$,
which is outside the stability region in (65).
The $M^2$ sum rule variable stability shown in their paper and
translated in
terms of the $\tau \equiv 1/M^2$ used in our paper
ranges between 0.04 and 0.13 GeV$^{-2}$,
in agreement with ours, but appears too
small compared with
 other channels studied until now within QSSR. This is because
the non-perturbative
terms do not play any essential role in the analysis.

Our results agree with the one indicated by the potential models
in (30). If we tentatively average the result in (63) with the previous
potential one in (30), we can deduce:
\beq
\FB|_{average} \simeq (2.94 \pm 0.21) f_\pi,
\label{3.15'''}
\eeq
which we consider to be our final estimate.
\subsection{The $\La(bcu)$ correlator}
\label{SS33}

Let us consider the baryonic current:
\beq
J=
r_1\,\left( u^t{\cal C}\gamma^5 c\right)b+
r_2\,\left( u^t{\cal C} c\right)\gamma^5 b+
r_3\,\left( u^t{\cal C} \gamma^5\gamma^\mu c\right)\gamma_\mu b
\label{3.17}
\eeq
which has
 the quantum numbers of the $\La(bcu)$; $r_1$, $r_2$ and $r_3$ are
arbitrary mixing parameters where, in terms of the $b$ parameter used in
Ref.~\cite{CHAB2}:
\beq
r_1=(5+b)/\;2\sqrt{6};\quad r_2=(1+5b)/\;2\sqrt{6};
 \quad r_3=(1-b)/\;2\sqrt{6}.
\label{3.18}
\eeq
The choice of operators in Ref.~\cite{CHAB4} is recovered
in the particular case where:
\beq
r_1=1;\quad r_2=k;\quad r_3=0.
\label{3.19}
\eeq
The associated two-point correlator is:
\beq
i\int d^4x\; \e^{ip\cdot x} \langle0|{\bf T}J(x)\overline{J}(0)|0\rangle
=\ps F_1 +F_2.
\label{3.20}
\eeq
The QCD expressions of the form factors $F_1$ and $F_2$ can be parametrized as:
\beq
F_i=F_i^{{\rm Pert}} +F_i^G+F_i^{{\rm Mix}},
\label{3.21}
\eeq
where:
\bea
\mbox{Im}\;F_2^{\rm Pert}(t)&=&{1\over128\pi^3 t}
\left\{
            (2 r_3^2+r_2^2-r_1^2)\,m_b\,
\left\{
6
\left[
            m_b^2 t^2+(m_b^4-2 m_b^2 m_c^2-m_c^4 )t \right.\right.\right.
\nonumber\\
&+& 2 \left.
                    m_b^2 m_c^4                                  \right]
\,{\cal L}_1-6t\left[
m_b^2\,t+(m_b^2-m_c^2)^2\right]\,{\cal L}_2\nonumber\\
&-&\left. \left[
          t^2+5(2m_b^2-m_c^2)t+ m_b^4-5m_b^2m_c^2-2m_c^4
              \right]\;\lambda_{bc}^{1/2}
                                                                   \right\}
\nonumber\\
 &-&\alto 2r_1 r_3\,m_c\left\{
6\left[m_c^2 t^2+(m_c^4-2 m_c^2 m_b^2-m_b^4 )t+2 m_c^2 m_b^4\right]
{\cal L}_1\right. \nonumber\\
&+& 6t\left[
m_c^2\,t+(m_c^2-m_b^2)^2 \right]{\cal L}_2\nonumber\\
&-&
\left.\left.
\left[
t^2+5(2m_c^2-m_b^2)t+ m_c^4-5m_c^2m_b^2-2m_b^4
\right]\;\lambda_{cb}\right\}
\right\}
\label{ZZ1}\\
\mbox{Im}\;F_2^\psi(t)&=&{\langle\overline{\psi}\psi\rangle\over8\pi t}
\lambda_{bc}^{1/2}
\left\{
-(r_1^2+r_2^2+4 r_3^2)m_b m_c
+r_1 r_3\; (m_b^2+m_c^2-t)
\right\}
\label{ZZ2}\\
\mbox{Im}\;F_2^G(t)&=&{\langle\alpha_s\, G^2\rangle
\over384\pi^2 t}
\left\{ \left[
2 {r_3^2\over m_b}\left( -2t+7m_b^2+2m_c^2\right)\right.\right.\nonumber\\
&+&  {r_1^2
-r_2^2\over m_b}\left(2t+5m_b^2-2m_c^2 \right) +
2 {r_1 r_3 \over m_c}\left( 2 t-2m_b^2-m_c^2\right)
\nonumber\\
&+& \left. 12 r_2 r_3\, m_c \phantom{{r_3^2\over m_b}}\kern-5.0mm
\right]\lambda_{bc}^{1/2}\nonumber\\
&+& 6\left[(r_2^2-r_1^2)\,m_b\,t+2r_3^2 m_b m_c^2-r_1 r_3 m_c t-r_2 r_3
m_c\,(t-2m_b^2)
\right]{\cal L}_1 \nonumber\\
&-& \left. 6t\left[ (r_2^2-r_1^2) m_b + (r_1+r_2)r_3 m_c
\right]\,{\cal L}_2
\phantom{{r_3^2\over m_b}}\kern-5.0mm \right\}
\label{ZZ3}\\
\mbox{Im}\;F_2^{\rm Mix}(t)&=&{M_0^2\langle
\overline{\psi}\psi\rangle \over64\pi t\, \lambda_{bc}^{3/2}}
\left\{2(r_1^2+r_2^2)\, m_b m_c \left[-t^3+t^2(m_b^2+3m_c^2)
\right.\right.\nonumber\\
&+&\left. t(m_b^2+m_c^2)(m_b^2-3m_c^2)-(m_b^2-m_c^2)^3\right]\nonumber\\
&+&       4r_3^2 m_b m_c \left[ -t^3+t^2(3m_b^2+m_c^2)\right. \nonumber\\
&+&\left.
 t(-3m_b^4-6m_b^2m_c^2+m_c^4)
+(m_b^2-m_c^2)^3\right]
\nonumber\\
&+& 2 r_1 r_3
\left[ t^4+t^3(-3m_b^2-2m_c^2)+3t^2m_b^2(m_b^2-m_c^2)\right. \nonumber\\
&+&\left. t(-m_b^6+4m_b^4 m_c^2+ 3m_b^2 m_c^4+2m_c^6)
+m_c^2(m_b^2-m_c^2)^3\right]\nonumber\\
&+& 2 r_2 r_3 \left[
 t^4+t^3(-4m_b^2-3m_c^2)+3t^2(2m_b^4+m_b^2m_c^2+m_c^4)\right. \nonumber\\
&-&\left.\left. t(m_b^2-m_c^2)(4m_b^4+ m_b^2 m_c^2-m_c^4)
+m_b^2(m_b^2-m_c^2)^3
\right]
\right\}
\label{3.22}\\
\mbox{Im}\;F_1^{\rm Pert}(t)&=&{1\over 512\pi^3 t^2}
\left\{
(r_1^2+r_2^2+4r_3^2)\left\{
12\left[t^2(m_b^4+m_c^4)-2m_b^4m_c^4\right]\,{\cal L}_1
\right.\right.\nonumber\\
&-&12t^2(
m_b^4-m_c^4)\,{\cal L}_2 \nonumber\\
&+& \left[ t^3-7t^2(m_b^2+m_c^2)+t(-7m_b^4+12m_b^2m_c^2-7m_c^4)\right.
\nonumber\\
&+&\left.\left.
m_b^6-7m_b^4m_c^2-7m_b^2m_c^4+m_c^6\right]\;\lambda_{bc}^{1/2}
\right\}  \nonumber\\
 &-& 4r_1 r_3\,m_b m_c\left\{ 12\left[
 t^2(m_b^2+m_c^2)-4tm_b^2m_c^2+2m_b^2m_c^2(m_b^2+m_c^2)\right]
{\cal L}_1 \right.\nonumber\\
&-& 12t^2(m_b^2-m_c^2)\,{\cal L}_2\nonumber\\
&-& \left. \left. 2\left[
2t^2+5t(m_b^2+m_c^2)-m_b^4-10m_b^2m_c^2-m_c^4
\right]\;\lambda_{cb}\right\}
\right\}\label{ZZ4}\\
\mbox{Im}\;
F_1^\psi(t)&=&{\langle\overline{\psi}\psi\rangle\over16\pi t^2}
\lambda_{bc}^{1/2}
\left\{
(2r_3^2+r_2^2- r_1^2)\, m_c (t+m_b^2-m_c^2)\right.\nonumber\\
&+&\left. 2 r_1 r_3\, m_b (m_b^2-m_c^2-t)
\right\}
\label{ZZ5}\\
\mbox{Im}\;F_1^G(t)&=&{\langle\alpha_s\, G^2\rangle\over768\pi^2 t^2}
\left\{ \left[
-4r_3^2 \left( t+3m_b^2\right)-
(r_2^2
+r_1^2) \left(t-3m_b^2+3m_c^2 \right) \right.\right.\nonumber\\
&+&
4{r_1 r_3 \over m_b m_c}\left( 2 t\,(m_b^2+m_c^2)-2m_b^4
-11m_b^2m_c^2-2m_c^4\right)
\nonumber\\
&-& \left. 36 r_2 r_3\, m_b m_c
\right]\lambda_{bc}^{1/2}\nonumber\\
&+& 12 m_b m_c \left[- 2 r_3^2\,m_b m_c +2 r_1 r_3 \left(
t-2m_b^2-3m_c^2 \right)
               \right. \nonumber\\
&+& \left. \left. 2 r_2 r_3 \left(
t-m_b^2-2m_c^2          \right)
   \right]   {\cal L}_1
          \right\}
\label{ZZ6}\\
\mbox{Im}\;F_1^{\rm Mix}(t)&=&{M_0^2\langle
\overline{\psi}\psi\rangle \over64\pi t^2\, \lambda_{bc}^{3/2}}
\left\{2(r_1^2-r_2^2)\,m_c \left[-t^4+t^3(2m_b^2+5m_c^2)
\right.\right.\nonumber\\
&-& t^2(2m_b^4
+3m_b^2m_c^2+9m_c^4)
+t(m_b^2-m_c^2)(2m_b^4-m_b^2m_c^2-7m_c^4)\nonumber\\
&-&\left. (m_b^2-m_c^2)^3(m_b^2-2m_c^2)\right]\nonumber\\
&+&       2r_3^2 m_c \left[ t^3(m_b^2-m_c^2)+t^2(-3m_b^4+4m_b^2
m_c^2+3m_c^4)\right. \nonumber\\
&+&\left.
 3t(m_b^2-m_c^2)(m_b^2+m_c^2)^2
-(m_b^2-m_c^2)^3(m_b^2+m_c^2)\right]
\nonumber\\
&+&  2r_1 r_3\,m_b
\left[ -t^4+t^3(5m_b^2+m_c^2)+t^2(-9m_b^4-4 m_b^2 m_c^2+m_c^4)\right.
 \nonumber\\
&+&\left. t(m_b^2-m_c^2)(7m_b^4+4m_b^2 m_c^2+m_c^4)
-2m_b^2(m_b^2-m_c^2)^3\right]\nonumber\\
&+& 2 r_2 r_3 m_b\left[
 -t^4+2 t^3(2m_b^2+m_c^2)-2t^2(3m_b^4+m_c^4)\right. \nonumber\\
&+&\left.\left. 2 t (2m_b^6-3m_b^4 m_c^2+m_c^6)
-(m_b^2-m_c^2)^4
\right]
\right\},
\label{3.23}
\eea
with:
\beq
\barr{lcllcl}
{\cal L}_1(t)\!\!\!&=&\!\!\!\displaystyle{{1\over2}\log{1+v\over1-v};}&
v\!\!\!&=&\!\!\!\displaystyle{\sqrt{1-{4m_b^2m_c^2\over (t-m_b^2-m_c^2)^2}}}\\
\lambda_{bc}^{1/2}\!\!\!&=&\!\!\!(t-m_b^2-m_c^2)\,v;&
{\cal
L}_2\!\!\!&=&\!\!\!\displaystyle{\log{(m_b^2+m_c^2)t+(m_b^2-m_c^2)
(\lambda_{bc}^{1/2}-m_b^2+m_c^2)
\over 2m_b m_c \,t}}.\\
 &=&\!\!\!\lambda^{1/2}(m_b^2,m_c^2,t)& & &
\ear
\label{3.24}
\eeq
The QCD expressions in Ref.~\cite{CHAB4} are recovered for the values of $r_i$
in~(\ref{3.19}). Those in Ref.~\cite{CHAB2} are obtained by
taking the value of
$r_i$ in~(\ref{3.18}), letting $m_c\to0$. This is a non-trivial check that we
now discuss in some detail. For the perturbative part one has to take into
account that:
\bea
{\cal L}_1&{\buildrel m_c\to0\over\longrightarrow}&{1\over2}\log{t\over
m_c^2}+ {1\over2}\log{(t-m_b^2)^2\over m_b^2 t}+\cdots
\nonumber\\
{\cal L}_2&{\buildrel m_c\to0\over\longrightarrow}&{1\over2}\log{t\over m_c^2}+
{1\over2}\log{m_b^2 (t-m_b^2)^2\over t^3}+\cdots
\label{3.25}
\\
\lambda_{bc}^{1/2}&{\buildrel m_c\to0\over\longrightarrow}&t-m_b^2+\cdots
\nonumber
\eea
For the quark condensate, one
must recall that when $m_c\to0$ the $c$-quark
must be allowed to condense. The easiest way to find this new
$c$-quark condensate contribution consists in
isolating the $1/m_c$ poles in the gluon condensate coefficients and using
the first term of the heavy-quark expansion in~(\ref{b}).
The $m_c$ pole parts are
\begin{eqnarray}
\mbox{Im}\, \left. F_2^G\right|_{\hbox{\scriptsize $m_c$-pole}}&=&
{r_1 r_2\over 96\pi^2 t} \left(
 t-m_b^2 \right)^2\langle \alpha_s G^2 \rangle {1\over m_c}\nonumber\\
&{\buildrel {\rm eq.(\protect\ref{b})} \over \longrightarrow}&
-{5-4b-b^2\over192\pi}\langle \overline{\psi}\psi\rangle
{(t-m_b^2)^2\over t}\nonumber\\
\mbox{Im} \, \left. F_1^G \right|_{\hbox{\scriptsize $m_c$-pole}}&=&
{r_1 r_3\over 96\pi^2 t^2}\,m_b \left(
t-m_b^2 \right)^2 \langle \alpha_s G^2 \rangle {1\over m_c}\nonumber\\
&{\buildrel {\rm eq.(\protect\ref{b})} \over \longrightarrow}&
-{5-4b-b^2\over192\pi}m_b\langle \overline{\psi}\psi\rangle
{(t-m_b^2)^2\over t^2}.
\label{3.26}
\end{eqnarray}
Adding these contributions to $\lim_{m_c\to0}$Im\,$ F_2^\psi$ and
$\lim_{m_c\to0}$Im\,$ F_2^\psi$, as can be read off~(\ref{3.22})
and~(\ref{3.23}), one gets agreement with the corresponding results in
Ref.~\cite{CHAB2}.

Similarly, to check the mixed condensate contributions one has to isolate
the $m_c\log{m_c}$ singularity of Im $\,F_1^G$,
Im $\,F_1^G$ and take into
account the first term of the heavy-quark expansion in~(\ref{c}).
The $m_c\log m_c$ singularities are
\begin{eqnarray}
\mbox{Im}\, \left. F_2^G \right|_{m_c\log m_c}
&=&-{\langle\alpha_s G^2\rangle(1-b)\over 768\pi^2 t}
\left\{m_b^2-6t+b(5m_b^2-6t)\right\}{m_c\over2}\log m_c^2\nonumber\\
&{\buildrel {\rm eq.(\protect\ref{c})} \over \longrightarrow}&
-{M_0^2\langle\overline{\psi}\psi\rangle(1-b)\over 768\pi t}
\left\{m_b^2-6t+b(5m_b^2-6t)\right\}\nonumber\\
\mbox{Im}\, \left. F_1^G \right|_{m_c\log m_c}
&=&{\langle\alpha_s G^2 \rangle(1-b)\over 768\pi^2 t^2}m_b
\left\{11m_b^2-6t+b(7m_b^2-6t)\right\}{m_c\over2}\log m_c^2\nonumber\\
&{\buildrel {\rm eq.(\protect\ref{c})} \over \longrightarrow}&
{M_0^2\langle\overline{\psi}\psi\rangle(1-b)\over 768\pi t^2}m_b
\left\{11m_b^2-6t+b(7m_b^2-6t)\right\}.
\label{3.27}
\end{eqnarray}
Adding these equations to $\lim_{m_c\to0}$Im\, $F_1^{\rm Mix}$,
$\lim_{m_c\to0}$Im\,$ F_2^{\rm Mix}$, one recovers the
corresponding expressions in Ref.~\cite{CHAB2}.

Finally, one must check the non-singular part of the gluon condensate
coefficients, i.e.
\bea
\mbox{Im}\,F_2^G|_{\hbox{\scriptsize non-sing}}&=&\mbox{Im}\,F_2^G-
\mbox{Im}\,F_2^G|_{\hbox{\scriptsize $m_c$-pole}}
-\mbox{Im}\,F_2^G|_{m_c\log m_c}\\
\mbox{Im}\,F_1^G|_{\hbox{\scriptsize non-sing}}&=&\mbox{Im}\,F_1^G-
\mbox{Im}\,F_1^G|_{\hbox{\scriptsize $m_c$-pole}}
-\mbox{Im}\,F_1^G|_{m_c\log m_c},
\eea
which should agree with Im $\,F_1^G$,
Im $\,F_2^G$ in Ref.~\cite{CHAB2}. This is the most difficult part,
for one must compute the $c$-quark condensate to order $m_c$ and use
again~(\ref{b}) to disentangle the misplaced quark-condensate contributions
{}from the genuine gluon condensate ones. The desired pieces are
\begin{eqnarray}
\mbox{Im}\,F_2^\psi&=&\dots+{11+2b-13b^2\over192\pi t}
\langle\overline{c}c\rangle m_c \,m_b(t-m_b^2) +O(m_c^2)\nonumber\\
&{\buildrel {\rm eq.(\protect\ref{b})} \over \longrightarrow}&
-{11+2b-13b^2\over2304\pi^2 t}
\langle\alpha_s G^2\rangle \,m_b(t-m_b^2)\nonumber\\
\mbox{Im}\,F_1^\psi&=&\dots+{5+2b+5b^2\over128\pi t^2}
\langle\overline{c}c\rangle m_c \,(t^2-m_b^4) +O(m_c^2)\nonumber\\
&{\buildrel {\rm eq.(\protect\ref{b})} \over \longrightarrow}&
-{5+2b+5b^2\over1536\pi^2 t^2}
\langle\alpha_s G^2\rangle\,(t^2-m_b^4).
\label{3.28}
\end{eqnarray}
Subtracting again these pieces from
$\lim_{m_c\to0}$ Im $\,F_2^G|_{\hbox{\scriptsize non-sing}}$
and from
$\lim_{m_c\to0}$Im $\,F_1^G|_{\hbox{\scriptsize non-sing}}$,
we obtain the corresponding
coefficients in Ref.~\cite{CHAB2}, as we should.

\subsection{The $\La(bcu)$ mass and coupling}
\label{SS34}

The $\La(bcu)$ contribution to the spectral function can be parametrized as:
\beq
\barr{lclcl}
{1\over\pi}\mbox{Im} F_1(t)&=&\left|Z_\La\right|^2\de(t-M_\La^2)&+&
\theta(t-tc)\times\mbox{\rm `QCD continuum'}\\
{1\over\pi}\mbox{Im} F_2(t)&=&M_\La \left|Z_\La\right|^2\de(t-M_\La^2)&+&
\theta(t-tc)\times\mbox{\rm `QCD continuum'}
\ear
\label{3.29}
\eeq
{}From the analogue of the sum rules in eqs.~(\ref{3.11})--(\ref{3.14}),
one can determine the residue $|Z_\La|$ and the $\La$-mass.
\begin{figure}[hb!]
 \centering
 \includegraphics[width=.7\textwidth]{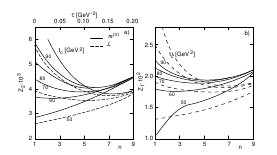}
 \caption{{\bf a)} $n$- and $\tau$-dependences of
the coupling $Z_\Lambda$ from
$F_1$ in (90),
for
different values of the continuum threshold $t_c$.
The continuous (dashed) lines come from the moments
(Laplace) sum rules analysis.
{\bf b)} The same as in {\bf a)} but from $F_2$.}
 \label{fig:fig3}
\end{figure}

The analysis for the residue is
shown in Fig.~3
for $b=-1/5$ (we have checked that
the result is insensitive to the value of
$b$ between $-1$ and $+1$ though the convergence of the OPE is bad for
$|b|\ge0.5$). As can be seen in this figure, the $\tau$ or $n$
stability is reached for $t_c \geq 60$ GeV$^2$, while the $t_c$ stability
starts at $t_c = 90 $GeV$^2$. We consider this range of values for our
optimal estimate. Then, we obtain from the $F_1$ and $F_2$ sum rules :
\beq
|Z_\La|^2 \simeq (4.0\sim 20.0)10^{-3}\, \mbox{GeV}^6,
\eeq
which is quite inaccurate as other QSSR estimates of the baryon
couplings in the heavy quark sector\cite{CHAB1}--\cite{CHAB4}.
For the estimate of the $\La$ mass, we use the ratios of sum rules.
However, these quantities do not present an
$n$/$\tau$ stability. We therefore fix the value of $n$/$\tau$ at the one where
$|Z_\La|$ is $\tau$-stable.
The $t_c$-dependence of the $\La$-mass is quite small, as shown
in the Fig.~4
%
%
and we fix it in the range corresponding to the optimal value of
the residue. By taking the largest range of the predictions
{}from the $F_1$ and $F_2$ ratios of moments and Laplace
sum rules, we deduce the value: $(6.86 \pm 0.26)$ GeV.
\begin{figure}[hb!]
 \centering
 \includegraphics[width=.5\textwidth]{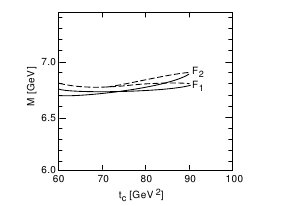}
 \caption{$t_c$-dependence of the $\Lambda$ mass from $F_1$ and $F_2$.}
 \label{fig:fig4}
\end{figure}

We add to the previous errors an error of about 100 MeV from $M_b$
and 10 MeV from the gluon condensate. Then, we deduce the final
estimate :
\beq
M_\La=(6.86\pm 0.28)\mbox{GeV}.
\label{3.30}
\eeq
in good agreement with the potential model estimate in~(\ref{final-baryons}).
This
value
is about $400\, \mbox{\rm MeV}$ higher than the previous result in
Ref.~\cite{CHAB2},
based on a particular choice of the operator.
\subsection{The ${\Xi^*_b}(bbu)$
and ${\Xi^*_c}(ccu)$ masses and  couplings}
For a comparison with the potential model results in Table 1, let us
remind the QSSR results obtained in \cite{CHAB4}:
\beq
M_{\Xi^*_c} \simeq (3.58 \pm 0.05)\mbox{GeV}  \; \; \; \;
M_{\Xi^*_b} \simeq (10.33\pm 1.09)\mbox{GeV}.
\eeq
These predictions agree quite well with the results in Table 1
with a similar accuracy for $\Xi^*_c$. The corresponding coupling
constants are:
\beq
|Z_{\Xi^*_c}^2\vert\simeq (3\sim 8)10^{-3}\, \mbox{GeV}^6, \; \; \;
|Z_{\Xi^*_b}^2 \vert\simeq (5\sim 23)10^{-3}\, \mbox{GeV}^6.
\eeq
The agreement of the different predictions between potential models and
QSSR calculations of the hadron masses is a good indication of the
convergence of the different theoretical estimates.

\section{Semileptonic decays of the $B_c$ mesons}
\label{S4}

\subsection{The procedures}
\label{SS41}

The first investigations of the three-point functions in the framework of
QCD spectral
sum rules have been performed in~\cite{IS82} for the form factor of the
pion. They have been
subsequently applied to semileptonic decays of heavy--light
mesons~\cite{BD} and heavy--heavy mesons~\cite{COLA}.
The first analysis of
the $t$-dependence of the semileptonic form factors was given by~\cite{BBD91b}.
We first shortly review the general sum rule technique for
the determination of
current matrix elements between heavy mesons.
Let $J_\mu$ be the weak current in
the quark sector:
\beq
J_\mu=:\Pb \ga_\mu (1-\ga^5)Q:,
\label{D.1}
\eeq
where $Q$ is the field for a heavy quark and $\psi$ for a light or heavy
one. We shall treat here the semileptonic decays of the
heavy--heavy meson
$B_c$, with the current
\beq
J_5=(m_b+m_c):\bar b (i\ga^5) c:.
\label{D.2}
\eeq
The decay product may be heavy--heavy ($\eta_c$, $J/\psi$) or
heavy--light ($B_s$,
$B_s^*$, $B$, $B^*$, $D$, $D^*$). For convenience we shall use here the method
for a pseudoscalar final state with $J_F=(m_\psi+m_Q):
\bar\psi(i\ga^5)Q:$. The starting
point for the SR analysis is the three-point function ($t=(p'-p)^2$):
\bea
\Pi_\mu(p,p')&=&i^2\int d^4x d^4y \e^{ip'\cdot x-ip\cdot y}
\langle0|{\bf T} J_F(x) J_\mu(0) J_{5}^\dagger (y)|0\rangle \nn\\
&=&i(p_\mu+p'_\mu)\Pi^+(p^2,p'^2,t)+i(p_\mu-p'_\mu)\Pi^-(p^2,p'^2,t).
\label{D.3}
\eea
In order to come to observables, we insert intermediate
states between the weak
and the hadronic current and obtain
\beq
\Pi_\mu(p,p')={\langle 0|J_F|H_F\rangle
\langle H_F|J_\mu|B_c\rangle
\langle B_c| J_{5}^\dagger|0\rangle
\over(p^2-M_{B_c}^2)(p'^2-M_{H_F}^2)}
+\hbox{\rm higher-state contributions}.
\label{D.4}
\eeq
$H_F$ is the lightest meson with the quantum numbers of $J_F(x)$, its mass is
$M_F$; $\langle H_F|J_\mu|B_c\rangle$ is the semileptonic decay matrix element
we are interested in. It can be decomposed as
\beq
\langle H_F|J_\mu|B_c\rangle=F_+(t)(p+p')_\mu+F_-(t)(p-p')_\mu .
\label{D.5}
\eeq
For semileptonic decays, only the form factor $F_+$ contributes as the
 contribution
of $F_-$ is proportional to the mass squared of the lepton. The factors
$\langle 0|J_F|H_F\rangle$ and $\langle B_c|J_{5}^\dagger|0\rangle$ are
proportional to the decay constants (see section~\ref{SS32}).
The contribution of
the higher states will be discussed later.

\begin{figure}[hb!]
 \centering
 \includegraphics[width=.5\textwidth]{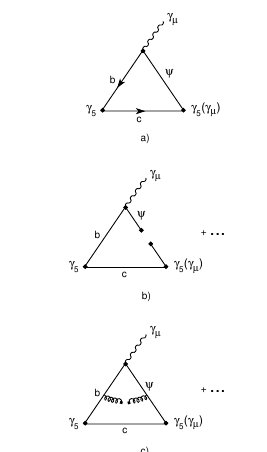}
 \caption{Different QCD contributions to the vertex functions: {\bf a)}
perturbative diagram, {\bf b)} light-quark condensate,
{\bf c)} gluon condensate.}
 \label{fig:fig5}
\end{figure}
As a next step,
 we evaluate the three-point function $\Pi_\mu$ in the framework
of QCD. In general one has to take into account perturbative (Fig.~5a)
and non-perturbative (e.g. Fig.~5b--c) contributions.
Since heavy quarks
do not condense and since even for the case of a light quark in the final
state the condensation of this quark (Fig.~5b) does not contribute to
the three-point sum rule, only the gluon condensate (Fig.~5c)
gives a non-perturbative
correction. This correction is, however, expected to be very small as has been
shown in~\cite{BALL}. Therefore, the ingredient that is dominant, by far,
is the
perturbative graph~(Fig.~5a).
The treatment of the higher power corrections
thus does not
play an essential role. They are taken into account by local
duality~\cite{SVZ}. If the perturbative contribution is represented by the
double dispersion relation:
\beq
\Pi_+(p^2,p'^2,t)=\int_{(m_Q+m_{Q'})^2}^\infty ds
\int_{(m_\psi+m_{Q'})^2}^\infty ds'
{\rho_+^{\rm pert}(s,s',t)\over(s-p^2)(s'-p'^2)},
\label{D.6}
\eeq
one assumes that for $p^2$, $p'^2$ sufficiently below the thresholds of $s$ and
$s'$ (say, $1$ GeV below)
the contribution of the higher states can be well
approximated by the perturbative contribution above certain thresholds $ t_c$,
$t_c'$.
%
%
We thus come to the sum rule:
\bea
\int_{(m_Q+m_{Q'})^2}^{t_c} ds \int_{(m_\psi+m_{Q'})^2}^{t_c'} ds'
{\rho_+^{\rm pert}(s,s',t)\over(s-p^2)(s'-p'^2)}+
\mbox{\rm ``non-pert. terms"} \nn\\
\approx
{\langle 0|J_F|H_F\rangle
\langle B_c| J_{5}^\dagger|0\rangle
F_+(t)
\over(p^2-M_{B_c}^2)(p'^2-M_{H_F}^2)}.
\label{D.8}
\eea

In order to suppress the dependence on the choice of
the ``continuum thresholds"
$t_c$, $t_c'$, the sum rule~(\ref{D.8}) is Borel- (Laplace-)
 transformed, yielding:
\bea
\int_{(m_Q+m_{Q'})^2}^{t_c} ds \int_{(m_\psi+m_{Q'})^2}^{t_c'} ds'
\rho_+(s,s',t) \e^{-s\tau} \e^{-s'\tau'}+\mbox{\rm ``non-pert. terms"}
\nn \\
=\e^{-M_F^2\tau'} \e^{-M_{B_c}^2\tau}
\langle 0|J_F|H_F\rangle
\langle B_c| J_{5}^\dagger|0\rangle
F_+(t).
\label{D.9}
\eea

In the next step, the matrix elements
$\langle 0|J_F|H_F\rangle$ and $\langle
B_c|J_{5}^\dagger|0\rangle$ are expressed through sum rules as done in
sections~\ref{SS31} and~\ref{SS32}. By choosing the parameters $\tau$ and
 $\tau'$ to be $1/2$ of the
corresponding parameter in the two-point sum rule, the exponential dependence
drops out, if we evaluate $F_+(t)$ from~(\ref{D.9}). Note that the sum
rule for the
two-point functions yield an expression for $|\langle 0|J_F|H_F\rangle|^2$ and
$|\langle
B_c|J_{5}^\dagger|0\rangle|^2$. We furthermore choose the continuum threshold
the same for the two- and three-point functions, i.e. $t_c$ for the $B_c$
 channel
and $t_c'$ for the $H_F$ channel. There is a very subtle point in the
$t$-dependence of the perturbative double spectral function. For $t<0$ there is
no problem in applying the Cutkosky rules
in order to determine $\rho_+(s,s',t)$
and the limits of integration. For $t>0$, which is the physical region for
decays, non-Landau-type singularities appear~\cite{BBD91b,BALL},
which make the
determination of the double spectral function very cumbersome.
For finite values
of $t_c$, $t_c'$, the non-Landau singularities do not contribute to the
sum rule~(\ref{D.9}) if $t$ is smaller than a certain value $t_{cr}$,
 which depends on
$t_c$ and $t_c'$, and hence the Cutkosky
rules may be applied in a straightforward way. For the determination
of the ratios, we extrapolate the $t$-dependence of that range to the
full range with a cubic extrapolation.

The continuum thresholds $t_c$ and $t'_c$ are parametrized by
\beq
t_c = (m_Q+m_{Q'}+E_c)^2 \ \ , \ \
t'_c = (m_Q+m_{Q'}+E'_c)^2 \ \ . \ \
\eeq
In many cases \cite{SN2}, \cite{ZAL}, \cite{CHAB4},
\cite{CHAB1}--\cite{CHAB3},
\cite{BBD91b}, \cite{BAG}, \cite{NEU}:
\beq
 E_c \simeq 1. \sim 2.~\mbox{GeV}
\eeq
yields optimal results for the
QSSR analysis. We shall use for definiteness the previous range in
our analysis.
In the evaluation, we do not take the (small) contribution from the
gluon condensate into account and we hence come to the following
sum rules :
\beq
F_+(t)={\e^{-M_F^2\tau'} \e^{-M_{B_c}^2\tau'} \over
\langle 0|J_F|H_F\rangle
\langle B_c| J_{5}^\dagger|0\rangle}
\int_{(m_Q+m_{Q'})^2}^{(m_Q+m_{Q'}+E_c)^2} ds
 \int_{(m_\psi+m_{Q'})^2}^{(m_Q+m_{Q'}+E'_c)^2} ds'
\rho_+(s,s',t) \e^{-s\tau} \e^{-s'\tau'}.
\label{D.11}
\eeq
For the case of a vector meson in the final state, the relevant
amplitudes are given by :
\bea
\langle H_F \ep^{(\la)}|J_\mu| B_c\rangle=
-iF_0^A(t) \ep^*_\mu +i F_+^A(t)\ep^{*(\la)}\cdot p\;(p+p')_\mu \nn\\
+2F_V(t)\ep_\mu^{\nu\rho\si}\ep^{*(\la)}_\nu p_\rho p'_\si +\dots
\label{D.12}
\eea
The amplitudes $F_+$ and $F_V$ receive their contributions from the
vector currents, while $F_0^A$ and $F_+^A$ do so
{}from the axial-vector one.
The relation between the scalar functions given in~(\ref{D.5})
and~({D.12})
and the ones used in Refs.~\cite{BWS 85,BBD91b} is
\beq
\begin{array}{lclclclc}
F_+&=&f_+&;&F_0^A&=&(M_{B_c}+M_f) A_1&;\\
F^A_+&=&\displaystyle{ {-A_2\over M_{B_c}+M_V} }&;&
F_V&=&\displaystyle{{V\over M_{B_c}+M_V}}&.
\end{array}
\label{D.13}
\eeq
For each of the amplitudes in~(\ref{D.12}), there is a sum
rule
like~(\ref{D.11}), with $\rho_+$ replaced by $\rho_0^A$, $\rho_+^A$ and
$\rho_V$ respectively. For completeness, we quote the relation of the
amplitudes to the decay rate.
In the case of the pseudoscalar final state,
we have :
\beq
{d\Gamma_+\over dt}={G_F^2|V_{Q\psi}|^2\over192\pi^3M_{B_c}^3}
\la^{3/2}(M_{B_c}^2,M_F^2,t) F_+^2(t),
\eeq
while for the vector final state :
\bea
{d\Gamma_+\over dt}&=&{G_F^2|V_{Q\psi}|^2\over192\pi^3M_{B_c}^3}
\la^{1/2}(M_{B_c}^2,M_F^2,t)\nn\\
&\times&
\left[
2(F_0^A)^2+\la F_V^2
+{1\over4M_F^2}
\left((M_{B_c}^2-M_F^2-t)F_0^A+\la F_+^A
\right)^2
\right],\nn\\
\la&=&\la(M_{B_c}^2,M_F^2,t).
\eea
\subsection{Results}
The principal results of the sum-rules evaluation of the form factors
in~(\ref{D.11}) are collected in Table~\ref{X.1}.
\begin{table}
    \begin{center}
\begin{tabular}{||c|c|c|c|c|c||}
   \hline
Channels & Reference & $f_+$ & $V$ & $A_2$ & $A_1$ \\
   \hline
$c\bar{c}$ & This
paper & $0.55\pm0.10$ & $0.48\pm0.07$ & $0.30\pm0.05$ & $0.30\
0.05$ \\
   \cline{2-6}
   & \protect{\cite{COLA}} & $0.20\pm0.01$ & $0.37\pm0.1$ & $0.27\pm0.03$ &
                         $0.28\pm0.01$ \\
   \hline
$b\bar{s}$ & This paper & $0.60\pm0.12$ & $1.6\pm0.3$ & $0.06\pm0.06$ &
 $0.40\pm 0.10$ \\
   \cline{2-6}
   & \protect{\cite{COLA}} & $0.30\pm0.05$ & $2.1\pm0.25$ & $0.39\pm0.05$ &
                         $0.35\pm 0.20$ \\
   \hline
$B\rightarrow D^{(*)}$
& \protect{\cite{BALL}} & $0.75\pm0.05$ &
                         $0.8\pm0.1$ & $0.68\pm0.08$ & $0.65\pm0.10$ \\
   \cline{2-6}
   & \protect{\cite{BWS 85}} & $0.69$ & $0.71$ & $0.69$ & $0.65$ \\
   \cline{2-6}
& \protect{[48]} & $0.62\pm 0.06$ &
                         $0.58\pm 0.03$ & $0.53\pm 0.09$ & $0.46\pm 0.02
                          $ \\
   \hline
\end{tabular}
\\[0.5cm]
\begin{tabular}{||l|c|c|c|c||}
   \hline
  & $B_c\rightarrow\eta_c $ & $B_c\rightarrow B_s   $ &
    $B_c\rightarrow B     $ & $B_c\rightarrow D     $ \\
  & $B_c\rightarrow J/\psi$ & $B_c\rightarrow B^*_s $ &
    $B_c\rightarrow B^*   $ & $B_c\rightarrow D^*   $ \\
\hline
$F_+(0)$ & $0.55\pm 0.10$& $0.60 \pm 0.12$ &$0.48 \pm 0.14$
  &$0.18 \pm 0.08$\\
\hline
$F_V(0)$
[GeV$^{-1}$] &$ 0.048\pm 0.007$ &$0.15 \pm 0.02$  &$0.11 \pm 0.02$
  &$0.02 \pm 0.01$\\
\hline
$F^A_+(0)$ [GeV$^{-1}$] &$-0.030 \pm 0.003$ &$-0.005 \pm 0.005$
 &$0.005 \pm 0.005$ &$ 0.010 \pm 0.010$ \\
\hline
$F^A_0(0)$ [GeV]
 &$3.0\pm 0.5$ &$3.3 \pm 0.7$ & $1.7 \pm 0.7$ &$0.8 \pm 0.4$\\
\hline
155--384 & (10--75)\ $10^3$ \\
\end{tabular}
    \end{center}
\caption{
\label{X.1} Comparison of semileptonic form factors for
different decays. We compare the dimensionless quantities $f_+$,
$A_1$, $A_2$, $V$ related to $F^A_0$, $F^A_+$ and $F_V$
through~(\protect{\ref{D.13}}).
}
\end{table}
The value with the lower
(resp. larger) modulus
corresponds to the value of the continuum energy $E_c= 1$GeV (resp.
$2$GeV).

\begin{table}
\begin{center}
\begin{tabular}{||c|c|c||}
   \hline
Channels & Reference & Rates in $10^{10} \mbox{\rm s}^{-1}$ \\
   \hline
$B_s l\nu$ & This paper & $0.35\pm0.10$ \\
   \cline{2-3}
   & \protect{\cite{COLA}} & $0.18$ \\
   \hline
$B^*_s l\nu$ & This paper & $0.35\pm0.10$ \\
   \cline{2-3}
   & \protect{\cite{COLA}} & $0.87$ \\
   \hline
$b\bar{s} l\nu$ & This paper & $ $ \\
   \cline{2-3}
   & \protect{\cite{COLA}} & \\
 \cline{2-3}
   & \protect{\cite{Eichten-Quigg-bc}} & $2.91$ \\
   \hline
$\eta_c l \nu$ & This paper & $0.27\pm0.07$ \\
   \cline{2-3}
   & \protect{\cite{COLA}} & $0.03$ \\
   \hline
$J/\psi l \nu$ & This paper & $0.32\pm0.08$ \\
   \cline{2-3}
   & \protect{\cite{COLA}} & $0.21$ \\
 \hline
$c\bar{c} l\nu$ & This paper & $ $ \\
   \cline{2-3}
   & \protect{\cite{COLA}} & $    $ \\
 \cline{2-3}
   & \protect{\cite{Eichten-Quigg-bc}} & $6.90$ \\
   \hline
\end{tabular}
\end{center}
\caption{\label{table4} Partial decay rates for $B_c$ and $B^*_c$
mesons}
\end{table}

\begin{figure}[hb!]
 \centering
 \includegraphics[width=.5\textwidth]{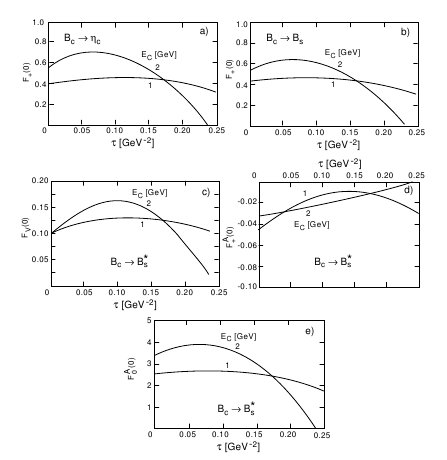}
 \caption{$\tau \simeq \tau'$-dependence
of the different form factors for $B_c$ semileptonic
decays at zero momentum transfer for
different values of the continuum threshold $E_c$ : {\bf a)}
$B_c \rar \eta_c$,
{\bf b)} $B_c \rar B_s$, {\bf c)--e)} $B_c \rar B^*_s$.}
 \label{fig:fig6}
\end{figure}
In Fig.~6,
we display the result of the form factors at $t=0$ as
function of $\tau $ (parameter of the initial state) $\simeq$ $\tau'$
(parameter of the final state). It shows a weak $E_c$-dependence
for $E_c$ in the range given in (103) while the $\tau$-stability is
roughly about one-half of the one from the two-point correlator (see
Fig. 3).

In Fig.~7,
\begin{figure}[hb!]
 \centering
 \includegraphics[width=.5\textwidth]{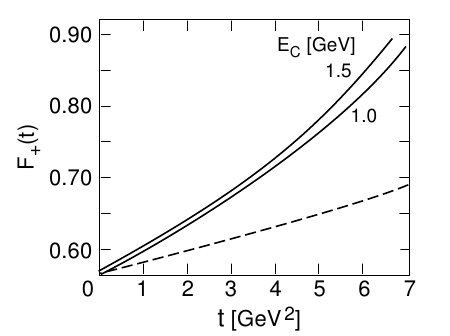}
 \caption{ $t$-dependence of the $B_c \rar \eta_c$ form factor:
 the continuous
lines are the QCD predictions using a polynomial fit; the dashed line is
the vector meson dominance prediction using a pole parametrization with
a $B^*_c$ mass of 6.33 GeV.}
 \label{fig:fig7}
\end{figure}
 we show the $t$-dependence of the form factor for the
semi-leptonic decay of $B_c$ into $\eta_c$ for $E_c$= 1 and 1.5 GeV. The
QSSR predictions with a polynomial fit
are represented by the continuous lines. The result
{}from the pole parametrization
\beq
F_+(t)={1\over 1-{t/M^2_{\rm pole}}}
\eeq
is given by the
 dashed line assuming a vector
dominance with a $B^*_c$ mass of $6.33\,$GeV. Our analysis indicates
that for large t-values the QCD prediction differs notably from
the pole parametrization within VDM.
 The same phenomena is observed in the other
channels as well.
For the $B_c$ into $J/\Psi$ semi-leptonic decay, we only quote the
fitted pole masses:
\bea
F_V:\quad M_{\rm pole}\simeq 4.08 \; \mbox{GeV}, \nonumber \\
F_+^A:\quad M_{\rm pole}\simeq 4.44 \; \mbox{GeV}, \nonumber \\
F_0^A:\quad M_{\rm pole}\simeq 4.62 \; \mbox{GeV},
\eea
needed for reproducing the QCD predictions.

\subsection{Discussions}

As mentioned above, the smallness of the non-perturbative corrections
is a particular feature of the
$\bar b c$ system.  The  analysis is rather an
application of local duality and the continuum model
than of the classical sum rules analysis as the stability of the
results versus the continuum threshold is only reached if one
assumes  that it is the same (however a natural choice) for the
two- and three-point functions.
 Nevertheless, we expect that
the ``physical'' results should lie in the range spanned by the
rather conservative choice of continuum thresholds, which corresponds
in different other channels to the optimal results from QSSR. The
choice of the continuum used in~\cite{DOM},\cite{COLA} does
not belong to this range and makes their results doubtful.

There is a considerable theoretical interest in the $t$-dependence of the
form factors for the heavy--heavy to heavy--heavy decays.
In~\cite{Ja90,JM92}, it has been shown in realistic models that the
$t$-dependence of the form factors of heavy--heavy mesons are not determined
by the lowest mass in the $t$-channel (vector meson dominance),
 but by the
size of the meson. This feature, which is obviously present in potential
models, is also visible in the sum rule analysis, as can be seen from
Fig.~7. The $t$-dependence is indeed much stronger than predicted
by vector meson dominance. Experimentally,
it would be important to
verify this deviation from a hadronic effective theory.

\section{Conclusions}
We have combined in this paper potential models and
QCD spectral sum rules for studying the properties of hadrons with charm
and beauty. We present in section 2 the results from potential models
with the emphasis on the accuracy of the models for predicting the
hadron masses. In section 3, we present the QCD spectral sum rules
estimates where we show that the values of the decay constants can come
out quite accurately once we use the meson masses from potential models
and once we understand better the Wilson coefficients in the Operator
Product Expansion of the correlators. Indeed, we show explicitly here
how the Wilson coefficients of the gluon condensates already contain
the ones of the heavy quark and heavy quark-gluon mixed condensates.
This point has been a source of confusion and uncertainties in the past.
Finally, we use in section 4 vertex sum rules in order to study the
form factors of the $B_c$ semileptonic decays. In particular, we show
that their $t$-dependence deviates notably from the one predicted by
vector meson dominance.
\section*{Acknowledgments}
We would like to thank Chris
Quigg for useful correspondence and Andr\'e Martin
for discussions. P.~G.~acknowledges gratefully a grant from the
Generalitat de Catalunya.
Part of this work has been done in the Institute of
theoretical physics of the University of Heidelberg, when S.N,
was an Alexander-Von-Humboldt Senior Fellow. This work has been
partially supported by CYCIT, project \#~AEN93-0520

\newpage


\begin{thebibliography}{10}

\bibitem{LHC}
A. De R\'ujula et al., {\it Proc. LHC Workshop}, Aachen 1990,
 CERN 90-10,{\bf Vo
 2},
p.201.

\bibitem{Eich-Fein}
E.~Eichten and F. Feinberg, Phys.~Rev.\ {\bf D23} (1981) 2724.

\bibitem{SN1}
S. Narison, Phys. Lett. {\bf B210} (1988) 238.

\bibitem{SVZ}M.A. Shifman, A.I. Vainshtein and V.I. Zakharov,
{ Nucl. Phys.} {\bf B147} (1979) 385, 448.

\bibitem{SN2}For a recent review, see e.g: S. Narison,
  Lecture Notes in Physics, {\bf Vol. 26}, {\it
QCD Spectral Sum Rules} (World Scientific, Singapore, 1989)
and references therein;

\bibitem{MAN}For reviews see e.g.
H. Georgi, {\it  Proceedings of TASI-91} (World Scientific, Singapore,
1991), edited by R.K. Ellis et al.;
N. Isgur and M. Wise,
{\it  Proceedings of Heavy Flavours} (World Scientific, Singapore,
1992), edited by A. Buras and M. Lindner;
T. Mannel, Talk given at the {\it 5th International
Symposium on Heavy Flavours}, Montreal, Canada, 6-10 June 1993,
CERN preprint TH-7052/93 (1993).

\bibitem{QR1}
C.~Quigg and J.L.~Rosner, Phys.\ Rep.\ {\bf 56} (1979) 167.

\bibitem{GM1}
H.~Grosse and A.~Martin, Phys.\ Rep.\ {\bf 60} (1979) 341.

\bibitem{Schlad}
A.~Martin, in {\it Proc.~Int.~Universit\"atswochen f\"ur Kernphysik},
  Schladming, Austria, 1986, ed.~H.~Latal and
  H.~Mittner (Springer Verlag,
  Berlin, 1987).

\bibitem{JMRrep}
J.-M.~Richard, Phys.\ Rep.\ {\bf 212} (1992) 1.

\bibitem{GMbook}
H.~Grosse and A.~Martin, book in preparation.

\bibitem{Thir}
See, e.g., W.~Thirring, {\it A Course in Mathematical Physics}, {\bf Vol.~3}
  (Springer Verlag, Berlin 1981).

\bibitem{BeMa}
R.A.~Bertlmann and A.~Martin, Nucl.~Phys. {\bf B168} (1980) 111.

\bibitem{PDG}
Particle Data Group,
 {\it Review of Particle Properties}, Phys.~Rev.\ {\bf D45}
  (1992) 1; {\bf D46} (1992) 5210 (E).

\bibitem{ccc2}
A.~Martin, in {\it Heavy Flavours and High Energy Collisions in the 1--100 TeV
  Range}, Proc.~Erice Workshop (1988), ed.~A.~Ali and L.~Cifarelli (Plenum,
  New York,~1989).

\bibitem{Gershtein}
S.S. Gershtein, V.V. Kiselev, A.K. Likhoded, S.R. Sla\-bo\-spit\-ski\v{\i}
  and A.V. Tka\-bla\-dze, Sov. J. Nucl. Phys. {\bf 48} (1988) 327.

\bibitem{Eichten-Quigg-bc}
E.~Eichten and C. Quigg, {\it Workshop on Beauty Physics at
  Colliders}, to appear in the Proceedings; Fermilab-pub-91/032-T (1994);
C. Quigg, Fermilab-Conf-93/265-T (1993).

\bibitem{God-Isg}
S. Godfrey and N. Isgur, Phys.~Rev.\ {\bf D32} (1985) 189.

\bibitem{LUSI}
M. Lusignoli and M. Masetti, Z. Phys. \ {\bf C51} (1991) 549.

\bibitem{Liebetc}
S.~Nussinov, Phys.~Rev.~Lett.\ {\bf 52} (1984) 966;\\ J.-M.~Richard and
  P.~Taxil, Phys.~Rev.~Lett.\ {\bf 54} (1985) 847;\\ E.~Lieb, Phys.~Rev.~Lett.\
  {\bf 54} (1985) 1987;\\ A.~Martin, J.-M.~Richard and P.~Taxil, Phys.~Lett.\
  {\bf 176B} (1986) 224.

\bibitem{Martin-bcs}
A. Martin, Phys.~Lett.\ {\bf B287} (1992) 251.

\bibitem{Post}
R.L.~Hall and H.R.~Post, Proc. R.~Phys.~Soc.\ {\bf 90} (1967) 381.

\bibitem{BMR2}
J.-L.~Basdevant, A.~Martin and J.-M.~Richard, Nucl.~Phys.\ {\bf B343} (1990)
  69.

\bibitem{FR1}
S.~Fleck and J.-M.~Richard, Progr. Theor.~Phys.\ {\bf 82} (1989) 760.

\bibitem{RiTa2}
J.-M.~Richard and P.~Taxil, Phys.~Lett.\ {\bf B128} (1983) 453.


\bibitem{Lambdab}
OPAL Collaboration, Contribution to the {\it 1993 EPS Conference},
Marseille, July 1993; we thank Fabienne Ledroit for this information.

\bibitem{CHAB4}
E. Bagan, M. Chabab and S. Narison, Phys. Lett. {\bf B306} (1993) 350.

\bibitem{HSF}
J.~Hiller, J.~Sucher and G.~Feinberg, Phys.~Rev.\ {\bf A18} (1978) 2399.

\bibitem{CohLip}
I.~Cohen and H.J.~Lipkin, Phys.~Lett.\ {\bf B106} (1981) 119.



\bibitem{BG1}D. Broadhurst and S.C. Generalis,
Open University preprint, OUT 4102-8 (1982) (unpublished); S.C. Generalis,
Ph.D Thesis, OUT 4102-13 (1982) (unpublished), \\
J. Phys. {\bf G16} (1990)
785, 367.

\bibitem{NOV} V.A. Novikov et al., Fortsch. Phys. {\bf 32} (1984) 585.


\bibitem{Jamin} M. Jamin and M. Munz, Z. Phys. {\bf C60} (1993) 569.

\bibitem{BrGe}D.~J.~Broadhurst and S.~C.~Generalis,
Phys.~Lett.~{\bf B139} (1984) 85; {\bf B142} (1984) 75;
{\bf B165} (1985) 175

\bibitem{SN3}
S. Narison, Z. Phys. {\bf C55} (1992) 671.

\bibitem{C_G^3}S.~N.~Nikolaev and A.~V.~Radyushkin, Nucl.~Phys.~{\bf B213}
(1983) 285.

\bibitem{Latorre} E.~Bagan, J.~I.~Latorre and P.~Pascual, Z.~Phys.~{\bf C32}
(1984) 75.

\bibitem{NSV}V.A. Novikov et al., { Nucl. Phys. } {\bf B237} (1984) 525.

\bibitem{SN4} S.~Narison, Phys.~Lett.~{\bf B198} (1987) 104 and {\bf B308}
(1993); Talk given at the
{\it Third $\tau$--Charm Factory
Workshop}, 1--6~June~1993, Marbella, Spain, CERN preprint
TH-7042/93 (1993).

\bibitem{ZAL}
S. Narison and M. Zalewski,
Phys. Lett. {\bf B320} (1994) 369.

\bibitem{DOM}
C.A. Dominguez, K. Schilcher and Y.L. Wu, Phys. Lett. {\bf B298} (1993)
190.

\bibitem{COLA}
P. Colangelo, G. Nardulli and N. Paver, Z. Phys. {\bf C57} (1993) 43.

\bibitem{CHAB1}E.~Bagan, M.~Chabab, H.~G.~Dosch and S.~Narison,
Phys.~Lett.~{\bf B287} (1992) 176.

\bibitem{CHAB2}
E. Bagan, M. Chabab, H.G. Dosch and S. Narison, Phys. Lett. {\bf B301} (1993)
 243.

\bibitem{CHAB3}
E. Bagan, M. Chabab, H.G. Dosch and S. Narison,
Phys. Lett. {\bf B278} (1992) 367.


\bibitem{IS82}
B.L. Ioffe and A.V. Smilga, Phys.
Lett. {\bf B114} (1982) 353; Nucl. Phys. {\bf
 B216}
(1983) 373.

\bibitem{BD}
 A. Ovchinnikov and V.A. Slobodenyuk,
{ Z. Phys.} {\bf C44} (1989) 433.
V.N. Baier and A.G. Grozin,
{ Z. Phys.} {\bf C47} (1990) 669.

\bibitem{BBD91b}P. Ball, V.M. Braun and H.G. Dosch,
 { Phys. Rev.} {\bf D44} (1991) 3567.

\bibitem{SN5}S. Narison, { Phys. Lett.} {\bf B283} (1992) 384.

\bibitem{BWS 85} M.~Bauer, B.~Stech and M.~Wirbel, Z.~Phys.~{\bf C29}
(1985) 637.

\bibitem{BAG}E. Bagan, P. Ball and P. Gosdzinsky,
{ Phys. Lett.} {\bf B301} (1993) 101.

\bibitem{NEU}M. Neubert, { Phys. Rev.} {\bf D45} (1992) 2451 and
{\bf D46} (1992) 1076.

\bibitem{BALL}P. Ball, Phys. Rev. {\bf D48} (1993) 3190.

\bibitem{Ja90}R.L. Jaffe, Phys. Lett. {\bf B245} (1990) 221.

\bibitem{JM92}R.L. Jaffe and P.F. Mende, Nucl. Phys. {\bf B369} (1992) 189.


\end{thebibliography}

\newpage

\section{Appendix}
\label{ap}

As anticipated in Sec.~\ref{SS31}, dispersion relations such as
\begin{equation}
C_{G^2}(q^2)={1\over\pi}\int_0^\infty{dt\,{\rm Im}\, C_{G^2}(t)\over
t-q^2} \label{Ab}
\end{equation}
require adding to~(\ref{3.3}) $\de$-functions and derivatives of
$\de$-functions
in order for them to be finite (and correct). The reason for
that should be clear by noting that~(\ref{3.3}) behaves as
$[t-(m_b+m_c)^2]^{-5/2}$ near theshold,
thus giving a divergent contribution to~(\ref{Ab}). The evaluation of
these extra terms can be rather cumbersome. Here we present a simpler
alternative modification of~(\ref{Ab}) which one can prove without much
effort. For the sake of simplicity, we illustrate the method with the $\GG$
contribution.
\noindent
Let us start by explicitly substituting~(\ref{3.3}) in~(\ref{Ab}):
\begin{equation}
C_{G^2}(q^2)=\!\!{1\over\pi}\!\!\int_{(m_b+m_c)^2}^\infty\!\!\!\!\!\!\!
dt
{ -\alpha_s  m_b m_c \,t\,\left( t\!-\!m_b^2\!\!-\!\!m_bm_c\!\!-\!\!
m_c^2\right)
\over2\,(t\!-\!q^2)\,[t\!-\!(m_b\!-\!m_c)^2]^{3/2}}\times
{[t\!-\!(m_b\!+\!m_c)^2]^{-5/2}}.
\label{A-c}
\end{equation}
Next, we separate the singular power of $t\!-\!(m_b\!+\!m_c)^2$, i.e.
the factor ${[t\!-\!(m_b\!+\!m_c)^2]^{-5/2}}$ in~(\ref{A-c}), from the
analytic portion and compute its Taylor series in powers of
$t\!-\!(m_b\!+\!m_c)^2$
up to order one.
Higher order terms are innecessary since they would
give a convergent contribution to~(\ref{A-c}) near theshold.
The desired Taylor
series is~($-q^2=Q^2>0$):
\begin{eqnarray}
&-&{\alpha_s \sqrt{m_b m_c}\over16[Q^2+(m_b+m_c)^2]}
\Bigg\{
(m_b+m_c)^2+{t-(m_b+m_c)^2\over8m_b m_c [Q^2+(m_b+m_c)^2]}\nonumber\\
&\times& \Big[
5 (m_b+m_c)^4+ Q^2
\left(
5m_b^2+18m_b m_c+5m_c^2
\right)
\Big]
\Bigg\}.\label{Ac}
\end{eqnarray}
Obviously, by subtracting eq.(\ref{Ac}) times $[t-(m_b+m_c)^2]^{-5/2}
$ from the integrand of~(\ref{A-c}) we obtain a result which is $O\{
[t-(m_b+m_c)^2]^{-1/2}\}$.
Thus,
this difference can be  integrated  as in~(\ref{Ab}). This is precisely
the modification of~(\ref{Ab}) that we are looking~for. So, we finally
have
\begin{eqnarray}
C_{G^2}(q^2)\!\!\!\!&=&\!\!\!\!{1\over\pi}\int_{(m_b+m_c)^2}^\infty
\!\Bigg\{\!\!
-{\alpha_s  m_b m_c \,t\,\left( t\!-\!m_b^2\!-\!m_bm_c\!-\!m_c^2\right)
\over2\,(Q^2\!+\!t)\;[t\!-\!(m_b\!-\!m_c)^2]^{3/2}}
\label{Ad}
\\
&-&\!\!{\alpha_s \sqrt{m_b m_c}}
\Bigg\{
{(m_b\!+\!m_c)^2[t\!-\!(m_b\!+\!m_c)^2]
\over16[Q^2+(m_b+m_c)^2]^2}-
{1\over16[Q^2\!+\!(m_b\!+\!m_c)^2]}
\nonumber\\
&\times& \Big[
(m_b\!+\!m_c)^2+{
(
5m_b^2\!+\!18m_b m_c\!+\!5m_c^2
)[t-(m_b+m_c)^2]\over8 m_b m_c}
\Big]
\Bigg\}
\Bigg\}\nonumber\\
&\times&{dt\over[t\!-\!(m_b\!+\!m_c)^2]^{5/2}}.
\nonumber
\end{eqnarray}
We have explicitely checked that~(\ref{Ad}) agrees with~\cite{SN2}.
An entirely analogous procedure can be followed to obtain the dispersion
relation for $C_{G^3}$.
One has
\def\S{\Sigma}
\def\SS{{\S^2}}
\begin{eqnarray}
C_{G^3}(q^2)&=&{1\over\pi}\int_\SS^\infty {dt\over[t-\SS]^{9/2}}
\Bigg\{  {\alpha_s m_b m_c\; t\over 6\;
(t+Q^2)\,[t-(m_b-m_c)^2]^{7/2}}
                                                       \nonumber\\
&\times&
\Big\{   3t^4
 -2(3m_b^2+2m_bm_c+3m_c^2)t^3  \nonumber\\
&+&  (5m_b^3m_c+18m_b^2m_c^2+5m_bm_c^3)\;t^2\nonumber\\
&+&   2(3m_b^6+m_b^5m_c-6m_b^4m_c^2-6m_b^3m_c^3-6m_b^2m_c^4+
m_bm_c^5+3m_c^6)\;t     \nonumber\\
&-&   3(m_b^8+m_b^7m_c-m_b^5m_c^3-2m_b^4m_c^4-m_b^3m_c^5+m_b
m_c^7+m_c^8)
     \Big\} \nonumber\\
&-&\alpha_s \sqrt{m_bm_c}
\left\{{-7\S^4\,(t-\SS)^3\over192\,(Q^2+\SS)^4}\right.\label{ZZZ1}\\
&+&
\left[{7\SS\,(t-\SS)^2\over192}+{A\,(t-\SS)^3\over 1536m_b m_c}\right]
{\SS\over (Q^2+\SS)^3}\nonumber\\
&+&\left[{-7\S^4\,(t-\SS)\over192}
-{\SS\,A\,(t-\SS)^2\over 1536m_b m_c}
+{B\,(t-\SS)^3\over24576m_b^2m_c^2}\right]{1\over(Q^2+\SS)^2}\nonumber\\
&+&\left.\left[{7\S^4\over192}+{\SS\,A\,(t-\SS)\over1536m_bm_c}
-{B\,(t-\SS)^2\over24576m_b^2m_c^2}-
{C\,(t-\SS)^3\over196608m_b^3m_c^3} \right]{1\over Q^2+\SS}
\right\}
\Bigg\},\nonumber
\end{eqnarray}
where we have introduced the notation:
\begin{eqnarray}
\Sigma&=&m_b+m_c\nonumber\\
A&=&51m_b^2+166m_b m_c+51m_c^2\nonumber\\
B&=&31m_b^4-836m_b^3m_c-1862m_b^2m_c^2-836m_bm_c^3+31m_c^4
\nonumber\\
C&=&277m_b^4+596m_b^3m_c-514m_b^2m_c^2+596m_bm_c^3+277m_c^4.
\label{ZZZ2}
\end{eqnarray}
The result is seen to agree with previous
calculations of the (real) part of~$C_{G^3}$ in the case
$m_b=m_c$~\cite{Latorre}.
Note that from~(\ref{Ad}) and~(\ref{ZZZ1}) it is straightforward to calculate
both the Borel- (Laplace-)
transform of~$C_{G^2}$ and~$C_{G^3}$ and their moments
since the dependence on $Q^2$
is through~$(Q^2+t)^{-n}$ or/and~$(Q^2+\SS)^{-n}$.
\end{document}